\documentclass[twocolumn, superscriptaddress,amsfont,amssymb,amsmath, showpacs,balancelastpage, nofootinbib]{revtex4-2}
\usepackage{graphicx,longtable,mathrsfs,color,array}
\usepackage[unicode=true,pdfusetitle,
bookmarks=true,bookmarksnumbered=true,bookmarksopen=true,bookmarksopenlevel=1,
breaklinks=false,pdfborder={0 0 0},backref=false,colorlinks=true]{hyperref}
\hypersetup{citecolor=blue,filecolor=blue,linkcolor=blue,urlcolor=blue}
\usepackage[usenames,dvipsnames,table]{xcolor}
\usepackage{amssymb,amsmath,mathtools,mathrsfs,enumitem}
\usepackage{epsfig,subfigure,placeins,float}
\usepackage{booktabs,multirow}
\usepackage{exscale,relsize}
\usepackage[normalem]{ulem}
\usepackage[T1]{fontenc}
\usepackage[utf8]{inputenc}
\usepackage{enumerate}
\usepackage{times, mathptmx}
\usepackage{tikz}
\usepackage{aas_macros}
\usepackage{bm}
\usetikzlibrary{arrows,positioning,decorations.markings,decorations.pathmorphing,calc}

\usepackage[capitalise]{cleveref}
\usepackage{diagbox}
\usepackage{hhline}

\newcommand{\be}{\begin{equation}}
\newcommand{\ee}{\end{equation}}

\newcommand{\comment}[1]{}

\newcommand{\mSt}{\{\mu_{\rm{today}},\Sigma_{\rm{today}}\}}
\newcommand{\mS}{\{\mu, \Sigma\}}
\newcommand{\mt}{\mu_{\rm{today}}}
\newcommand{\St}{\Sigma_{\rm{today}}}

\newcommand{\wzwa}{\{w_0, w_a\}}
\newcommand{\cBM}{\{c_B, c_M\}}
\newcommand{\ITk}{\text{ISW-}T\kappa}
\newcommand{\ITg}{\text{ISW-}Tg}
\newcommand{\txtp}{3\times 2\text{pt}}
\newcommand{\lcdm}{\Lambda\text{CDM}}
\newcolumntype{C}[1]{>{\centering\let\newline\\\arraybackslash\hspace{0pt}}m{#1}}

\newcommand{\togglered}[1]{\textcolor{black}{#1}}

\definecolor{hyperref}{RGB}{026,028,087}
\def\gsim{ \lower .75ex \hbox{$\sim$} \llap{\raise .27ex \hbox{$>$}} }
\def\lsim{ \lower .75ex \hbox{$\sim$} \llap{\raise .27ex \hbox{$<$}} }

\graphicspath{{./figures/}}
\allowdisplaybreaks

\tikzstyle{vecArrow} = [thick, decoration={markings,mark=at position
1 with {\arrow[semithick]{open triangle 60}}},
double distance=1.4pt, shorten >= 5.5pt,
preaction = {decorate},
postaction = {draw,line width=1.4pt, white,shorten >= 4.5pt}]

\begin{document}
\title{Constraining dark energy with complementary probes of large-scale structure}

\author{Neel Shah}
\affiliation{Institute of Cosmology \& Gravitation, University of Portsmouth, Portsmouth, PO1 3FX, U.K.}
\affiliation{Department of Physics \& Astronomy, University College London, London, WC1E 6BT, U.K}

\author{Kazuya Koyama}
\affiliation{Institute of Cosmology \& Gravitation, University of Portsmouth, Portsmouth, PO1 3FX, U.K.}
\affiliation{Kavli IPMU (WPI), UTIAS, The University of Tokyo, \\ Kashiwa, Chiba 277-8583, Japan}
\affiliation{Yukawa Institute for Theoretical Physics, Kyoto University, \\ Kyoto 606-8502, Japan}

\author{Johannes Noller}
\affiliation{Department of Physics \& Astronomy, University College London, London, WC1E 6BT, U.K}
\affiliation{Institute of Cosmology \& Gravitation, University of Portsmouth, Portsmouth, PO1 3FX, U.K.}

\author{Lanyang Yi}
\affiliation{National Astronomical Observatories, Chinese Academy of Sciences,\\
Beijing 100101, P.R.China}
\affiliation{School of Astronomy and Space Sciences, University of Chinese Academy of Sciences,\\
Beijing 100049, P.R.China}
\affiliation{Institute of Cosmology \& Gravitation, University of Portsmouth, Portsmouth, PO1 3FX, U.K.}

\begin{abstract}
To observationally pin down the nature of dark energy, it is essential to consistently model cosmological perturbations in the presence of dark energy alongside the background expansion and constrain this joint theory space with a large array of complementary probes. Here, we achieve this by constraining a model in the Effective Field Theory of Dark Energy (EFTofDE) framework by supplementing probes of the expansion history with several probes of large-scale structure: redshift space distortions (RSD) from DESI DR1, $3\times2$pt measurements from DES Y3, and the Integrated Sachs-Wolfe effect from cross-correlating CMB temperature anisotropies with galaxy number counts or CMB lensing. We demonstrate the complementarity of different probes which leads to strong improvements on constraints on DE perturbations. For our most constraining dataset combination that supplements CMB+BAO+SNe probes with DESI DR1 RSD, DES Y3 $3\times2$pt and ISW cross-correlations between CMB temperature and galaxy counts, we find an improvement in the Figure of Merit (FoM) for the DE perturbation parameters $\{c_B, c_M\}$ by a factor of 2.69. We show the phenomenological implications of these constraints by mapping them to the present-day values of the phenomenological functions $\{\mu(z), \Sigma(z)\}$, where we see an FoM improvement by a factor of 3.37. We find a significant interdependence between the posteriors of $\{w_0, w_a\}$ and $\{c_B, c_M\}$, caused by the theoretical prior imposed by the gradient stability condition within the EFTofDE framework. Finally, we compute the significance of deviation from $\Lambda$CDM for the EFTofDE model when constrained with CMB+BAO+SNe datasets, finding it to be at 2.9$\sigma$. This significance is nontrivially similar to the significance for the $w_0w_a$CDM model for the same dataset combination which we find to be 3.1$\sigma$.
\end{abstract}

\date{\today}
\maketitle

\tableofcontents

\section{Introduction} \label{sec:intro}

Identifying the nature of dark energy (DE) has been one of the core pursuits in cosmology ever since the discovery of the accelerated expansion of the universe \cite{Riess:1998cb,Perlmutter:1998np}, and this pursuit has been renewed due to recent observational hints for dynamical DE \cite{Rubin:2023jdq,DES:2024jxu,DESI:2024mwx,DESI:2024kob,DES:2025upx,DESI:2025zgx,DESI:2025fii,DES:2025sig,Hoyt:2026fve}.
The main phenomenological avenue for looking for dynamical DE signatures has been the background cosmic expansion history. The most well-established probes of the cosmic background expansion are the cosmic microwave background (CMB), baryon acoustic oscillations (BAO), and Type IA supernovae (SNe), with current state-of-the-art constraints coming from e.g. \cite{Planck:2018vyg,ACT:2025fju,SPT-3G:2025bzu,DES:2024jxu,DESI:2024mwx,DESI:2025zgx,Brout:2022vxf,Rubin:2023jdq,DES:2025upx,DES:2025sig,Hoyt:2026fve}. Constraints on dynamical DE signatures on the evolution of cosmic large-scale structure (LSS) are less explored and have scope to be improved by employing already existing LSS observations. Dynamical DE signatures on the evolution of cosmological perturbations contain further information about the microphysics of DE and are closely related to signatures of modified gravity. There is a wide variety of probes that constrain cosmological perturbations: CMB lensing, redshift space distortions (RSD), weak gravitational lensing, the Integrated Sachs-Wolfe (ISW) effect, peculiar velocities, and more (see e.g. \cite{Planck:2018lbu,ACT:2023kun,SPT-3G:2024atg,DESI:2024jxi,Wright:2025xka,DES:2026fyc,Carron:2022eum,Qin:2025ggt,Lai:2025xkf,Qin:2025rwz} for recent cosmological constraints).

A particularly useful framework for constraining the presence of dynamical dark energy systematically is the effective field theory of dark energy (EFTofDE) \cite{Gubitosi:2012hu,Bloomfield:2012ff,Gleyzes:2013ooa,Bellini:2014fua,Gleyzes:2014rba} (essentially equivalent to the Horndeski class of modified gravity theories on a cosmological background \cite{Horndeski:1974wa,Charmousis:2011bf,Deffayet:2011gz,Kobayashi:2011nu}) and related constructions (e.g. \cite{Lagos:2016gep,Lagos:2017hdr}), that parametrise a wide class of theoretically well-motivated models of dark energy -- typically with a focus on linear perturbations, although non-linear extensions of these approaches exist, see e.g. \cite{Kimura:2011dc,Takushima:2015iha,Cusin:2017mzw,Brando:2023fzu,Amendola:2025xka,Sirera:2026klo}. This complements more phenomenological approaches that e.g. parametrise modified gravity effects by explicitly modifying the Poisson-like equations relating the matter density perturbations to the Newtonian potential (which affects clustering of nonrelativistic matter) or the Weyl potential (which affects gravitational lensing of photons) \cite{Caldwell:2007cw,Amendola:2007rr,Hu:2007pj,Jain:2007yk,Bertschinger:2008zb,Zhao:2008bn,Pogosian:2010tj,Zhao:2010dz,Silvestri:2013ne}. These phenomenological modifications can likewise be computed in EFTofDE approaches, so the EFTofDE can be used as a theoretical prior in deriving such phenomenological parametrisations as well (e.g. \cite{Perenon:2016blf,Peirone:2017ywi,Espejo:2018hxa,Shah:2025vnt}). We will primarily use the EFTofDE framework here -- specifically we will work in the $\alpha$-basis pioneered by \cite{Bellini:2014fua} -- see \cite{Hu:2013twa,Raveri:2014cka,Bellini:2015xja,Gleyzes:2015rua,Zumalacarregui:2016pph,Alonso:2016suf,Arai:2017hxj,Kreisch:2017uet,Noller:2018wyv,Frusciante:2018jzw,Reischke:2018ooh,Mancini:2018qtb,Brando:2019xbv,Arjona:2019rfn,Raveri:2019mxg,Perenon:2019dpc,Frusciante:2019xia,SpurioMancini:2019rxy,Bonilla:2019mbm,Baker:2020apq,Joudaki:2020shz,Noller:2020lav,Noller:2020afd,Mastrogiovanni:2020gua,Gsponer:2021obj,Traykova:2021hbr,Castello:2023zjr,Seraille:2024beb,Chudaykin:2024gol,Toda:2024fgv,Taule:2024bot,Ishak:2024jhs,Colangeli:2025bnb,Lu:2025gki,Euclid:2025tpw,Shah:2025vnt,Lu:2025sjg,Stolzner:2025vet,Li:2026sbr} for related constraints. 
Results using this basis can straightforwardly be mapped to other related approaches. We will also link this to the phenomenological functions $\mS$ described later. 

Recently, analyses focusing on deriving EFTofDE constraints by combining multiple recent LSS datasets (on top of the standard CMB+BAO+SNe probes of the background expansion) have started to appear: e.g. \cite{Seraille:2024beb}, which obtains constraints combining CMB-galaxy cross-correlations probing the ISW effect and RSD measurements from BOSS and 6dF; \cite{Ishak:2024jhs}, which combines DESI DR1 full shape measurements with $\txtp$ data from DES Y3; \cite{Stolzner:2025vet}, which combines cosmic shear from KiDS-Legacy with RSD from eBOSS; \cite{Lu:2025sjg}, which combines CMB lensing and ISW-lensing cross-correlations from Planck PR4, full-shape information from BOSS, and photometric clustering data from DESI Legacy Imaging Survey DR9; and \cite{Li:2026sbr}, which combines gravitational potential decay rates measured from DESI photometric clustering and Planck, the lensing-inferred amplitude of matter perturbations from cross-correlation of Planck anisotropies with galaxy surveys, and growth rate measurements from DESI DR1 clustering and the DESI DR1 peculiar velocity survey.

In this work, we derive constraints on the EFTofDE perturbations and background expansion by supplementing the CMB+BAO+SNe combination with RSD measurements from DESI DR1 \cite{DESI:2025fii}, weak lensing $\txtp$ measurements from DES Y3 \cite{DES:2022ccp}, and ISW measurements from CMB-galaxy cross-correlations \cite{Stolzner:2017ged} and CMB temperature-lensing cross-correlations \cite{Carron:2022eyg,Carron:2022eum}. We present constraints from various combinations of these datasets to demonstrate the complementarity of these probes. Furthermore, we demonstrate and quantify how the posteriors of the background expansion and DE perturbations affect other due to the underlying physics of the gradient stability condition in the EFTofDE, and how this affects the observational significance of dynamical DE over a cosmological constant.
\\
   
{\bf Outline}: 
This paper is organised as follows. We introduce the underlying theoretical EFTofDE framework we are using, as well as its relation to the phenomenological $\mS$ functions in \cref{sec:EFT}. This is followed by a detailed description of the datasets and likelihoods we are using in \cref{sec:data}. In \cref{sec:combining_probes_constraints} we then present constraints obtained from various combinations of probes, discuss their complementarity and what these constraints mean for DE phenomenology. In \cref{sec:background_perts} we then focus specifically on the interplay between constraints on the DE background ($\{w_0,w_a\}$) and perturbation ($\cBM$) parameters, and evaluate the observational significance of deviation from $\lcdm$ in the EFTofDE model.
We conclude in \cref{sec:conclusions}, before collecting additional details in the appendices.
Throughout this paper we use natural units with $c = M_{pl} = 1$. We use dots to denote a derivative with respect to cosmic time.

\section{The EFT of Dark Energy} \label{sec:EFT}

The Effective Field of Dark Energy (EFTofDE) is a widely used approach to describe a general Lagrangian for dark energy, unifying several classes of DE models where the role of DE is effectively played by a single scalar field \cite{Gubitosi:2012hu,Bloomfield:2012ff,Gleyzes:2013ooa,Bellini:2014fua,Gleyzes:2014rba}. This approach enables the construction of the most general Lagrangian of the extra scalar degree of freedom compatible with the symmetries of the FLRW cosmological background up to a given order in derivatives/fields. In accordance with the symmetries of FLRW, the coefficients of the operators appearing in the EFTofDE Lagrangian do not depend on space to preserve spatial symmetry, and are free functions of time due to the broken time translation invariance of the FLRW background.

There has been significant progress in the past decade on developing a physically intuitive basis for the free functions of the EFTofDE \cite{Bloomfield:2012ff,Gubitosi:2012hu,Bellini:2014fua,Lombriser:2018olq}, and on constraining a variety of aspects of DE and modified gravity from astrophysical and cosmological observations. We choose to work in the $\alpha$-basis \cite{Bellini:2014fua,Gleyzes:2014rba}, and focus on the three free functions in the EFTofDE that are the most promisingly constrained by cosmological probes: the DE equation of state $w(z)$, the braiding $\alpha_B(z)$ describing the kinetic mixing between the scalar field and the metric, and the running of the Planck mass $\alpha_M(z)$ describing the time variation of the effective Planck mass squared seen by tensor perturbations, $M^2$.\footnote{This means we choose to investigate models where the speed of gravitational waves is the same as the speed of light. This is partially motivates by the near-simultaneous observations of GW170817 and GRB 170817A \cite{TheLIGOScientific:2017qsa,Goldstein:2017mmi,Savchenko:2017ffs,LIGOScientific:2017zic,GBM:2017lvd} and what this implies for cosmological models when the resulting constraint on the gravitational wave speed is directly ported to cosmological scales \cite{Creminelli:2017sry,Ezquiaga:2017ekz,Baker:2017hug}, but see \cite{deRham:2018red} for why this involves non-trivial assumptions and \cite{Harry:2022zey,LISACosmologyWorkingGroup:2022wjo,Baker:2022eiz,Sirera:2023pbs,Atkins:2024nvl} for possible ways to test these assumptions observationally. Furthermore, we choose a fiducial behaviour for the kineticity function $\alpha_K$ (see the text below \cref{cs2}), which is known to only affect dynamics beyond leading order in the quasi-static approximation \cite{Bellini:2014fua}. See \cite{Noller:2013wca,Sawicki:2015zya,Bellini:2019syt,Pace:2020qpj} for discussions of how and when this approximation is accurate.} \togglered{An important feature of the $\alpha$ basis is the explicit separation of the background and perturbations in the choice of time-varying degrees of freedom: The background expansion history is entirely determined by $w(z)$ and has no dependence on $\{\alpha_B(z), \alpha_M(z)\}$.}

Our main focus in this work is to study the complementarity of multiple probes and the interplay between background expansion and structure growth. We therefore adopt simple parametrisations of these functions that are widely used in the literature. We parametrise $w(z)$ with the CPL form \cite{Chevallier:2000qy,Linder:2002et}
\begin{equation}
w(a) = w_0 + w_a(1-a).
\end{equation}
We parametrise both $\alpha_B(z)$ and $\alpha_M(z)$ as \cite{Bellini:2014fua}
\begin{equation}
\alpha_B = c_B\Omega_{DE}(z),\quad \alpha_M(z) = c_M\Omega_{DE}(z),
\end{equation}
where $\Omega_{DE}(z)$ is the ratio of the energy density in DE to the critical density. This parametrisation for $\{\alpha_B(z), \alpha_M(z)\}$ is motivated by its simplicity of having a very small number of free parameters and by the assumption that DE and modified gravity are purely late-universe effects, such that gravity should return to GR at high redshifts\footnote{For the same reason, the effective Planck mass $M^2$ is assumed to be unity at early times. $\alpha_M$ describes its running via $\alpha_M(a) = d\ ln(M^2)/d\ ln(a)$.}.

An important physical quantity is the sound speed of the scalar field perturbations. It is determined from the EFTofDE functions by the expression

\begin{gather}\label{cs2}
c_s^2(z) =
\dfrac{1}{D}\Big\{(2 - \alpha_B)\left(\dfrac{1}{2}\alpha_B + \alpha_M - \dfrac{\dot{H}}{H^2}\right) 
\\  \nonumber 
- \dfrac{3(\rho_{tot} + p_{tot})}{H^2M^2} 
+ \dfrac{\dot{\alpha_B}}{H}\Big\}\ ,
\end{gather}
where we have suppressed time-dependences on the RHS for clarity, $D \equiv \alpha_K + \frac{3}{2}\alpha_B^2$ with $\alpha_K$ describing the kinetic term of the scalar field and $\rho_{tot}+p_{tot}$ is the sum over all non-DE components (essentially equal to $\rho_m$ for standard CDM at late times). $D >0$ is required for the absence of ghost instabilities. Since $\alpha_K$ is unconstrained from subhorizon measurements, we fix it to $\alpha_K = 0.1\Omega_{DE}$ to ensure that $D >0$ is always satisfied. To avoid spurious numerical instabilities due to oscillations in the equations of motion near radiation domination, we add a small constant (0.01) to $\alpha_K(z)$ called the "kineticity\_safe" parameter in hi\_class.

The phenomenological effects of the EFTofDE on linear cosmological perturbations are described by the relation between the gravitational potentials and the matter density perturbations, which are two Poisson-like equations: \cite{Caldwell:2007cw,Amendola:2007rr,Hu:2007pj,Jain:2007yk,Bertschinger:2008zb,Zhao:2008bn,Pogosian:2010tj,Zhao:2010dz,Silvestri:2013ne}
\begin{gather}\label{Poisson_mu}
\dfrac{k^2}{a^2}\Phi = -4\pi G\mu(z)\rho_m\delta_m\ ,\\ \label{Poisson_Sigma}
\dfrac{k^2}{a^2}(\Phi+\Psi) = -8\pi G\Sigma(z)\rho_m\delta_m\ .
\end{gather}
Here, $\mu(z)$ describes the modification to the evolution of the Newtonian potential $\Phi$ which affects the trajectories of non-relativistic matter, and $\Sigma(z)$ describes the modification to the evolution of the Weyl potential $\frac{1}{2}(\Phi + \Psi)$ which affects trajectories of light. Under the quasi-static approximations and assuming that the effective mass of the scalar field is light, we can assume that  $\mu(z)$ and $\Sigma(z)$ are scale independent and they 
can be expressed in terms of the EFTofDE parameters as \cite{Bellini:2014fua}
\begin{gather} \label{mu_qsa}
\mu(z) = \dfrac{1}{M^2}\left(1 + \dfrac{(\alpha_B + 2\alpha_M)^2}{2Dc_s^2}\right)\ ,\\ \label{Sigma_qsa}
\Sigma(z) = \dfrac{1}{M^2}\left(1 + \dfrac{(\alpha_B + \alpha_M)(\alpha_B + 2\alpha_M)}{2Dc_s^2}\right)\ .
\end{gather}

\section{Datasets and methodology} \label{sec:data}

In this section, we list all the complementary probes and corresponding datasets that we use to constrain modified gravity. We then explain how we modify our analysis choices to ensure that we don't combine correlated probes in the absence of a model for their covariance. We also explain how we compare the constraining power of different likelihood combinations on $\cBM$, which is non-trivial due to their posteriors being highly non-Gaussian.

\begin{figure*}[t]
    \centering
    \includegraphics[width=0.9\textwidth]{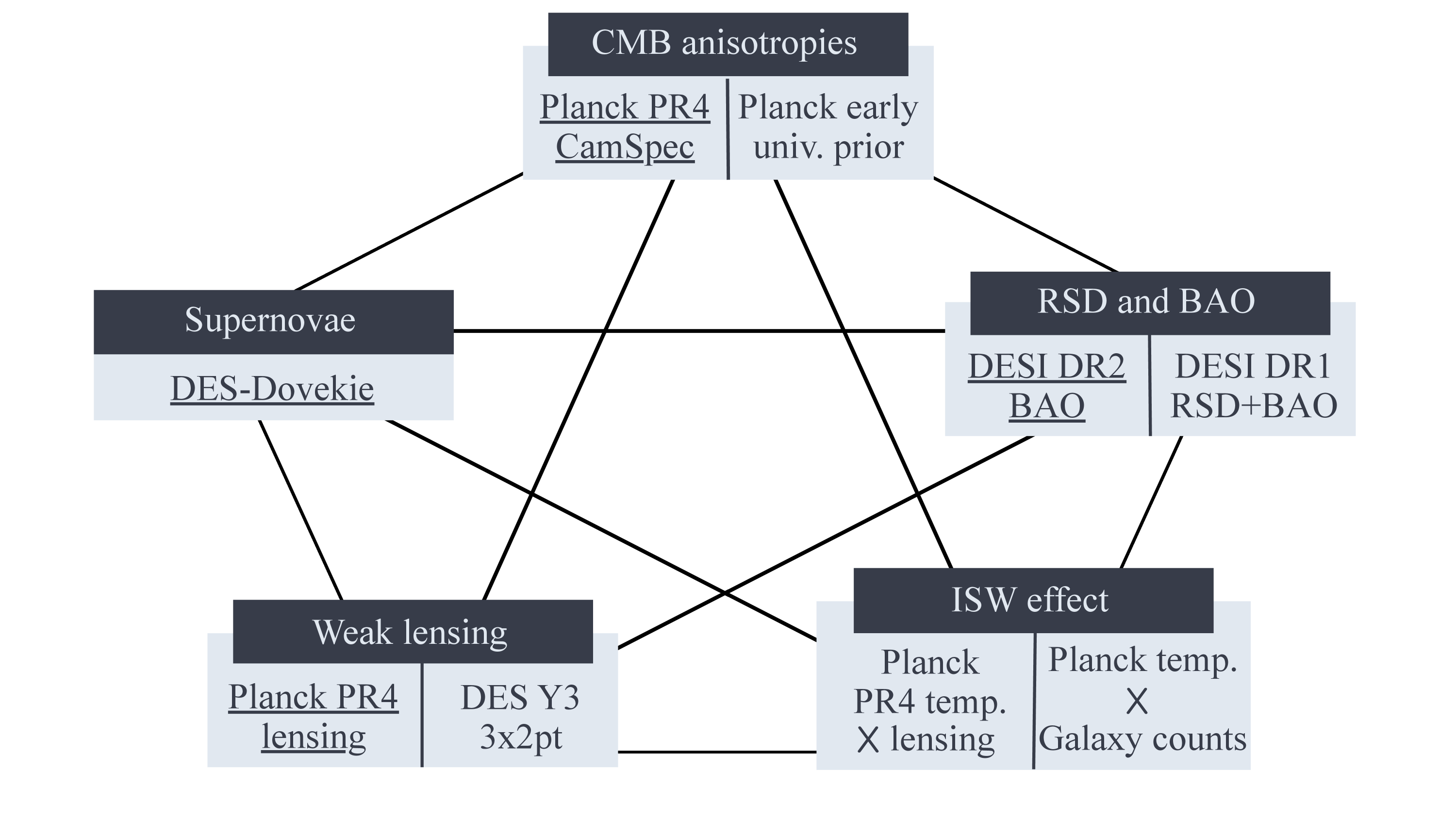}
    \caption{A graph showing all the likelihoods used in this work and their various possible combinations. We split our likelihoods into five blocks each representing a particular cosmological observable. Likelihoods in the same block probe very similar observables on overlapping regions of the universe, thus their covariance can be non-negligible, in which case combining them directly (without modelling the covariance) is unsound. Likelihoods probing different observables exhibit negligible correlations in their data vectors and systematics, and can therefore be combined without accounting for covariance; we use lines connecting the different blocks to represent all possible likelihood combinations. The underlined likelihoods are part of our minimal combination of probes that constrains EFTofDE perturbations $\cBM$ to $\mathcal{O}(1)$. The list of likelihoods used in this work are: Planck PR4 CamSpec \cite{Rosenberg:2022sdy,Planck:2018lbu,Planck:2019nip}, Planck early-universe prior \cite{Lemos:2023xhs}, DESI DR2 BAO \cite{DESI:2025zgx}, DESI DR1 RSD+BAO \cite{DESI:2024jxi}, Planck PR4 ISW-lensing \cite{Carron:2022eum,Carron:2022eyg}, Planck $\times$ galaxy surveys \cite{Stolzner:2017ged,Seraille:2024beb}, Planck PR4 lensing \cite{Carron:2022eyg}, DES Y3 $\txtp$ \cite{DES:2018gui,DES:2021rex,DES:2021wwk,DES:2022ccp}, and DES-Dovekie \cite{DES:2025sig,DES:2024jxu}.}
    \label{fig:probes_and_datasets}
\end{figure*}

\subsection{Likelihoods}

\begin{itemize}

\item{\bf CMB:} Observations of the cosmic microwave background (CMB) are sensitive to effects of late-time dark energy and modified gravity mainly through the late Integrated Sachs-Wolfe (ISW) effect governed by the time evolution of the gravitational potentials due to dark energy, and the gravitational lensing of CMB photons carried out by large scale structure in the late universe. CMB data is also crucial for constraining the remaining cosmological parameters apart from those related to dark energy. We use the CMB primary anisotropies (CMB-A) and lensing reconstruction (CMB-L). For the primary anisotropies, for high multipoles we use the CamSpec likelihood built on the Planck PR4 NPIPE maps \cite{Rosenberg:2022sdy}. The PR4 CamSpec likelihood has lower noise in the maps, and importantly, a lower significance of the lensing ($A_L$) anomaly compared to the official Planck 2018 likelihoods. In this work, we thus make the choice of fixing $A_L = 1$, noting that a preference for $A_L > 1$ is coincident with stronger preferences for modified gravity in a variety of modified gravity models when $A_L$ is fixed to unity, as shown by \cite{Pogosian:2021mcs,Specogna:2024euz,Shah:2025vnt}. For low multipoles, we use the official Planck 2018 likelihoods \cite{Planck:2018lbu,Planck:2019nip}. We also use the reconstruction of the CMB lensing potential using the Planck PR4 NPIPE maps \cite{Carron:2022eyg}, which exhibits an increase in constraining power compared to CMB lensing from Planck PR3 due to less noise in the PR4 maps as well improvements in the lensing analysis pipeline.

In \cref{sec:stability_prior}, we employ the CMB as a background-only probe of the late universe, like the BAO and SNe datasets. We do this with the help of the early universe prior derived from Planck PR4 maps, as obtained by the authors of \cite{Lemos:2023xhs}. This is a Gaussian prior on parameters affecting the early universe obtained by marginalising over late-universe information in the CMB.

\item{\bf RSD and BAO:} Galaxy redshift surveys measure baryon acoustic oscillations (BAO), which constrains the cosmic expansion history via measurements of the transverse and radial BAO distance scales, $D_M(z)/r_d$ and $D_H(z)/r_d$, as a function of redshift. They also measure redshift space distortions (RSD) from full-shape measurements of the galaxy power spectrum, which constrain the growth rate of structure at various redshifts. We use RSD and BAO measurements from the DESI collaboration. When we use the BAO data without RSD data, we use the DESI DR2 likelihood \cite{DESI:2025zgx}. When we include the RSD data, which are only publicly available for DESI DR1 at the time of writing, in order to account for the covariance between RSD and BAO measurements we switch back to DESI DR1 \cite{DESI:2024jxi}, and we refer to the combination of RSD and BAO measurements with a consistent modelling of their cross-covariance as RSD+BAO. We marginalise over the shape information captured by the ShapeFit parameters in the DESI full-shape measurements, and only use the $f\sigma_{s8}$ measurements in our analysis (where $s$ is the ratio of the sound horizon in the current cosmology to that in the reference cosmology of DESI and $\sigma_{s8}$ is the amplitude of the matter power spectrum computed at the scale $R = s \cdot 8h^{-1}$Mpc \cite{Brieden:2021edu,Lai:2024bpl}). Since the analytical expression for the growth rate $f$ valid in $\Lambda$CDM is not valid in EFTofDE, we perform an accurate numerical computation using the time derivative of the matter power spectrum amplitude $\sigma_{s8}$.\footnote{Technically, it is the power spectrum of only the clustering species: CDM + baryons i.e. $P_{cb}(k)$, that is used, rather than the total matter power spectrum $P_m(k)$ which also includes neutrinos. This is because scale-independence of growth and linear bias are better approximations for the power spectrum CDM + baryons \cite{Castorina:2013wga}. Nevertheless, we have checked that switching between $P_{cb}(k)$ and $P_m(k)$ gives a negligible difference in our predictions of $f\sigma_{s8}$.} As we show in \cref{tab:eft_constraints} and explain in \cref{sec:combining_probes_constraints}, the difference in constraining power between DESI DR2 and DR1 BAO is not substantial within EFTofDE when combined with probes of dark energy perturbations, and the posteriors for $\{w_0, w_a\}$ are very similar for the two BAO likelihoods.

\item{\bf $\bm{\txtp}$:} This refers to joint measurements (performed by photometric galaxy surveys) of cosmic shear, galaxy clustering and galaxy-galaxy lensing two-point correlation functions, which probe the growth of structure by constraining the gravitational lensing of the source galaxies' shapes by lens galaxies, the distribution of lens galaxies, and their cross-correlation. We use $\txtp$ measurements with the covariance of the three probes from three years of observations by the DES collaboration (DES Y3) \cite{DES:2018gui,DES:2021rex,DES:2021wwk,DES:2022ccp}. We use the linear scale cuts as provided by the collaboration obtained by the methods explained in \cite{DES:2022ccp} since non-linear modelling requires additional assumptions about higher-order interactions \cite{Kimura:2011td,Takushima:2015iha,Cusin:2017mzw}. The public DES Y3 likelihoods are interfaced with the CosmoSIS sampling code \cite{Zuntz:2014csq} and rely on some $\Lambda$CDM-like assumptions. We use the version of the likelihoods modified to work with Cobaya and to account for modified structure growth and lensing without any $\Lambda$CDM approximations, which was earlier used in \cite{Wang:2024uvw}.

\item{\bf ISW effect:} The detection significance of the ISW effect in the autocorrelation of the CMB temperature anisotropies is severely limited due to cosmic variance and the fact that the effect is smeared over a wide range of redshifts. The ISW signal increases significantly in measurements of the cross-correlation of CMB temperature anisotropies with a late-universe probe of structure, such as galaxy positions or CMB lensing, measured on wide angular scales. We use two ISW likelihoods in this work, which use two different ways to detect the ISW signal:

\begin{itemize}

\item{$\ITg$:} We use the ISW likelihood from \cite{Seraille:2024beb} -- an EFTofDE-adapted version of the ISW likelihood in \cite{Stolzner:2017ged} for $\Lambda{}$CDM. This employs measurements of the ISW effect from the cross-correlation of CMB temperature anisotropies with galaxy number counts from a number of galaxy catalogs: 2MPZ \cite{Bilicki:2013sza}, WISE $\times$ SuperCOSMOS \cite{Bilicki:2016irk}, SDSS-DR12 photo-z \cite{Beck_2016}, SDSS-DR6 QSO \cite{Richards:2008eq}, and NVSS \cite{Condon:1998iy}. The likelihood also incorporates autocorrelations of the galaxy number counts to break the degeneracy between galaxy bias and the ISW signal. Furthermore, we employ a simplification of the bias modelling justified in \cite{Seraille:2024beb}: namely we fix the redshift evolution of bias across all the bins of the same survey (such that $b_g(z) = b_{\rm survey}(1 + z)$ for each survey), thus reducing the number of bias parameters from one per survey bin to one per survey.

\item{$\ITk$:} We use the ISW signal measured from the lensing-ISW bispectrum from the cross-correlation of the Planck PR4 lensing quadratic estimator and temperature anisotropies \cite{Carron:2022eum,Carron:2022eyg}.\footnote{We refer to temperature-lensing cross-correlations as $C_\ell^{T\kappa}$ where $\kappa$ represents lensing convergence, instead of $C_\ell^{T\phi}$ in the original references in order to avoid confusion with the gravitational potentials or the scalar field in the EFTofDE.}

\end{itemize}

\item{\bf SNe:} Measuring the apparent luminosity of Type-IA supernovae over a range of redshifts provides constraints on the cosmic expansion history. We use the DES-Dovekie likelihood \cite{DES:2025sig}, which is a reanalysis of the DES 5-Year supernovae observations \cite{DES:2024jxu}. We note that recent developments in supernova modelling have substantially increased the consistency of expansion history constraints from different supernova collaborations \cite{Hoyt:2026fve}.

\end{itemize}

\subsection{Likelihood combinations}\label{sec:likelihood_combinations}

\cref{fig:probes_and_datasets} illustrates the mutual dependence of the various complementary probes and likelihoods that we use. Generally, different likelihoods for the same cosmological probe are expected to be correlated, while likelihoods of different probes can be considered to be mutually independent.

\begin{itemize}

\item{Minimal combination:} As our minimal combination of probes to constrain the EFTofDE background and perturbations, we combine the CMB anisotropies, CMB lensing, BAO, and SNe likelihoods. This combination, underlined in \cref{fig:probes_and_datasets}, provides strong constraints on all cosmological parameters not directly describing dark energy, while placing $\mathcal{O}(1)$ constraints on the perturbation parameters $\cBM$, and strong constraints on the background expansion parameters $\wzwa$, as the combination of CMB, BAO and SNe constrains the cosmic expansion history over a range of redshifts. We refer to this combination as "minimal" in the sense that although for constraints on the background expansion parameters $\wzwa$, this set of probes is already state-of-the-art, for the constraints on the parameters $\cBM$ describing DE perturbations, there is much room for improving the constraints by adding further probes of structure as we demonstrate in \cref{sec:combining_probes_constraints}. Throughout this paper, we refer to the constraints on DE obtained from this dataset combination as the minimal constraints.

\item{Adding RSD}: The DESI RSD and BAO measurements are both inferred from galaxies observed by the same survey and so are correlated. In data combinations with only BAO data, we use the DESI DR2 BAO likelihood, and in data combinations with both RSD and BAO data we use the joint measurements of RSD and BAO from DESI DR1, in order to account for the cross-covariance between RSD and BAO measurements.

\item{Adding $\txtp$:} The DES Y3 $\txtp$ likelihood incorporates measurements of cosmic shear, galaxy clustering, and galaxy-galaxy lensing. At linear order in perturbation theory, cosmic shear and the lensing convergence $\kappa$ are completely determined from each other, and so the $\txtp$ likelihood has non-trivial correlations with CMB lensing and $\ITk$ (as they also overlap in the redshift ranges and sky regions probed). In the absence of a covariance matrix between the corresponding observations, we thus cannot combine $\txtp$ with CMB lensing or $\ITk$, and when we add it to the minimal combination we are required to remove CMB lensing.

\item{Adding ISW:} The $\ITg$ and $\ITk$ likelihoods are naturally not independent
since both are based on extracting the ISW effect from the same CMB temperature anisotropies through cross-correlation. Since we do not have a model of their covariance, we cannot combine them in an analysis. The $\ITg$ likelihood is largely independent of any of our other likelihoods, and can thus be safely combined with them neglecting the covariance. The $\ITk$ likelihood, however, uses the same lensing potential reconstruction as the CMB lensing likelihood. The authors of \cite{Carron:2022eum} take the covariance of the ISW signal with CMB lensing into account, and it is part of the likelihood evaluation, which allows us to combine $\ITk$ with our minimal combination which includes CMB lensing without any extra modelling.

\end{itemize}

When it comes to combining our various likelihoods that include galaxy clustering: namely DES $\txtp$, $\ITg$, and DESI RSD, we can safely treat each of these as mutually independent. DESI RSD is independent of the other two likelihoods because it is a growth rate measurement which contains spectroscopic information about galaxy clustering along the line of sight, while the other two likelihoods contain information about transverse galaxy clustering obtained from photometric galaxy catalogs. We expect the covariance between $\txtp$ and $\ITg$ to be negligible due to the limited overlap of the survey footprints, the differences in redshift distributions, and the different angular scales probed by the ISW and $\txtp$ analyses.

We use the Cobaya code \cite{Torrado:2020dgo,2019ascl.soft10019T} for MCMC sampling and deriving posteriors of the parameters. We adopt $R - 1 < 0.03$ as the convergence criterion for our chains. We use the GetDist code \cite{Lewis:2019xzd} to compute convergence criteria, plot marginalised posteriors, and derive confidence limits on parameters.

\subsection{Quantifying and comparing constraining power for non-Gaussian posteriors}\label{sec:quantifying_constraining_power}

\begin{figure*}[!t]
    \centering
    \begin{subfigure}
        \centering
        \includegraphics[width=0.4\linewidth]{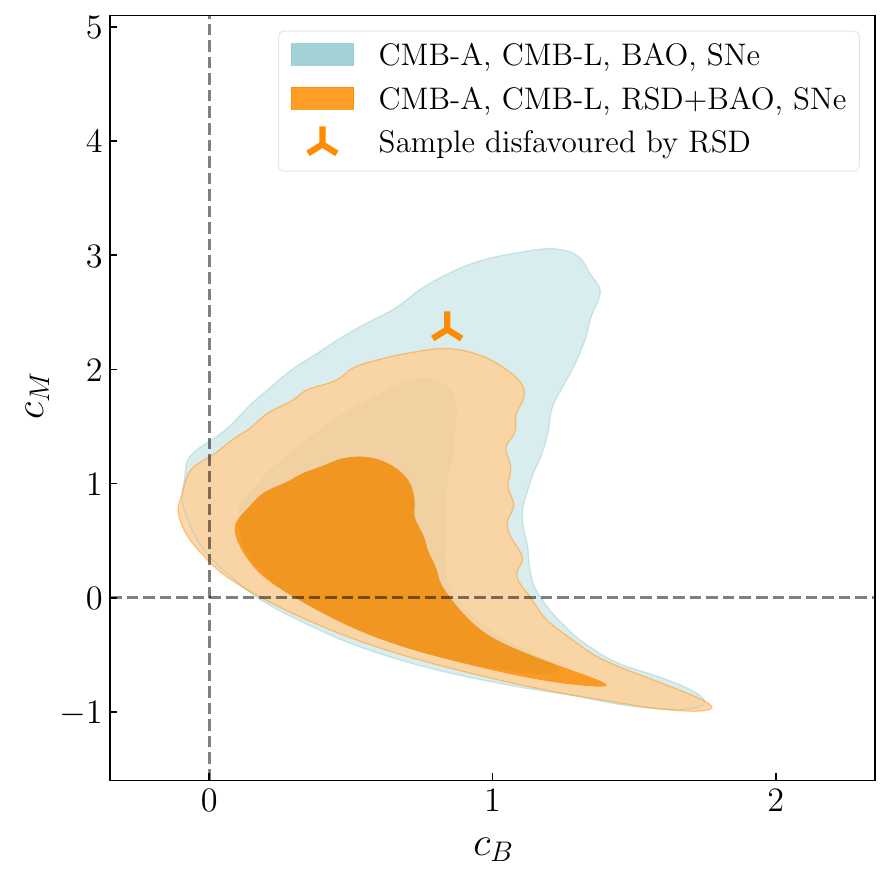}    
    \end{subfigure}
    \begin{subfigure}
        \centering
        \includegraphics[width=0.4\linewidth]{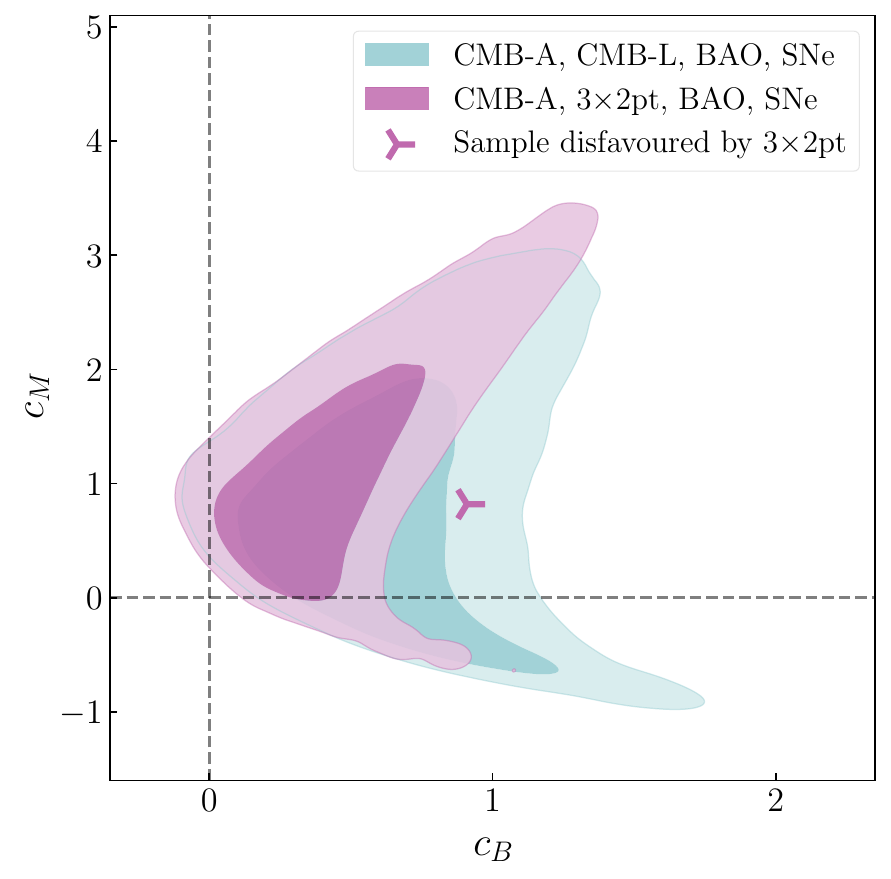}
    \end{subfigure}

    \begin{subfigure}
        \centering
        \includegraphics[width=0.4\linewidth]{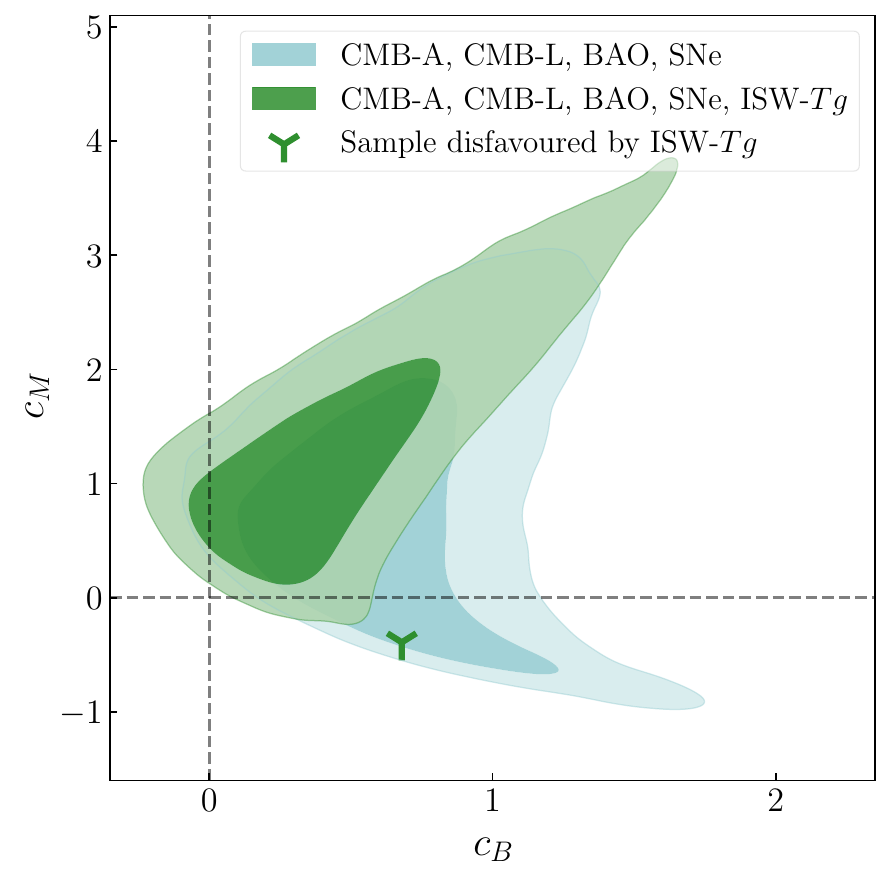}
    \end{subfigure}
    \begin{subfigure}
        \centering
        \includegraphics[width=0.4\linewidth]{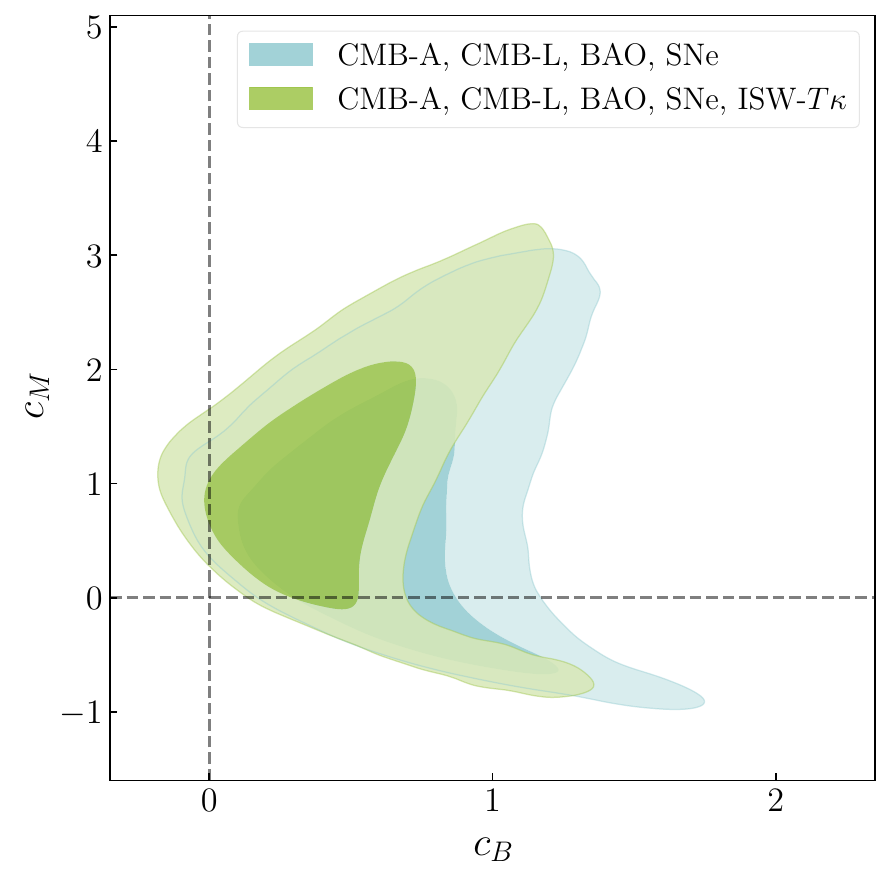}
    \end{subfigure}
    \caption{The impact on the constraints on DE perturbations in the EFTofDE on adding an extra probe to our minimal combination of probes which comprises of CMB anisotropies (CMB-A) and CMB lensing (CMB-L) from Planck PR4 CamSpec, baryon acoustic oscillations (BAO) from DESI DR2, and Type IA supernovae (SNe) from DES Y5. \textbf{Upper left:} Adding DESI DR1 RSD. Note that when adding this dataset, we replace DESI DR2 BAO in our minimal combination with DR1 BAO in order to incorporate the publicly available covariance between DESI DR1 RSD and BAO measurements. \textbf{Upper right:} Adding DES Y3 $\txtp$. Note that when adding this dataset, we remove CMB lensing from our minimal combination as the covariance between $\txtp$ and CMB lensing is non-negligible and unmodelled. \textbf{Lower left:} Adding the CMB $\times$ galaxies ISW likelihood ($\ITg$). \textbf{Lower right:} Adding the Planck PR4 ISW likelihood ($\ITk$). In the RSD, $\txtp$ and $\ITg$ plots, we also highlight $\cBM$ values of a particular sampled point from the posterior which is within the 95\% credible region for the minimal combination but disfavoured on adding the extra probe; these samples are used to demonstrate the phenomenological regimes constrained by these probes in \cref{sec:pheno_constraints}.}
    \label{fig:adding_one_probe}
\end{figure*}

In this work, we aim to understand and quantify how different probes of structure in the universe contribute to constraints on DE perturbations, how adding more probes improves the constraining power, and how different probes improve constraining power in complementary directions in the $\cBM$ parameter space. These quantifications are non-trivial because the posteriors for $\cBM$ are highly non-Gaussian as can be seen from \crefrange{fig:adding_one_probe}{fig:adding_three_probes}. This non-Gaussianity is a consequence of the fact that the Poisson equations \cref{Poisson_mu,Poisson_Sigma}, which are more directly connected to the observables, have a highly non-linear dependence on $\cBM$ and a far more complex time dependence than $\propto \Omega_{DE}$ as a result.

Due to this non-Gaussianity, traditional measures of posterior volume such as standard deviations of 1D marginalised posteriors or the traditional definition of the Figure of Merit (FoM) as being proportional to $(\det\bm{C})^{-1/2}$ where $\bm{C}$ is the posterior covariance of $\cBM$ are difficult to interpret. This expression for the FoM was originally given only for $\wzwa$ \cite{Albrecht:2006um} whose posteriors in standard analyses are Gaussian to a good approximation, and it doesn't have a helpful interpretation for non-Gaussian posteriors. One consequence of the non-Gaussianity is that the GR limit $(c_B, c_M) = (0, 0)$ is outside the 95\% credible region for most of the posteriors in \crefrange{fig:adding_one_probe}{fig:adding_three_probes}, but the individual parameters $c_B$ and $c_M$ appear more consistent with GR from most of these posteriors. In MCMC-sampled posteriors, the 68\% credible intervals are the most robustly sampled, while the 95\% and higher credible intervals are progressively more subject to numerical noise in the sampling. For these reasons, we use the following two metrics derived from the volume of the 68\% credible regions of our 1D and 2D posteriors of $\cBM$ to quantify constraining power of our probes:

\begin{itemize}
\item{Posterior width of 1D posteriors:} We define the posterior width of $c_B$ or $c_M$ as the difference between the upper and lower 68\% limits of the respective parameter. Note that all our 1D posteriors for $c_B$ or $c_M$ are unimodal and so the 68\% credible regions are indeed a single continuous interval, however the definition of the posterior width as the volume of the 68\% credible region is straightforward to generalise to multimodal posteriors.

\item{Figure of Merit (FoM) of 2D posteriors:} We define our FoM for $\cBM$ as the inverse area of the 68\% credible region of the 2D marginalised posterior of $\cBM$, which is a modification to the definition of the FoM for $\wzwa$ defined in \cite{Albrecht:2006um} where it was defined as the inverse area of the 95\% region. We choose the 68\% regions as they are are more robust to sampling noise than the 95\% regions. Since we are only interested in comparisons between FoMs of different dataset combinations/physical models, we normalise our FoM definition to set the FoM of the $\cBM$ posterior of the minimal dataset combination to unity, as shown in the first row of \cref{tab:eft_constraints}.

Note that we compute our FoMs numerically, as the analytical relation FoM $\propto (\det \bm{C})^{-1/2}$ where $\bm{C}$ is the covariance matrix of the two parameters does not hold for non-Gaussian posteriors. To compute the FoM of our posteriors normalised with respect to the FoM of the minimal combination, we simply take the ratio
\begin{gather}\label{fom_definition}
\text{FoM} \equiv \dfrac{\text{Area of 68\% contour in minimal combination}}{\text{Area of 68\% contour}}, \nonumber\\ 
\end{gather}
where the areas are simply computed by counting cells over a grid in the $\cBM$ space which lie inside the 68\% credible region of their 2D posterior. The accuracy of this method with respect to grid resolution is easy to ensure by checking that FoMs are negligibly affected by increasing the grid resolution. Another potential concern is smoothing artefacts due to the fact that the posteriors plotted in the countours are smoothed versions of the true posteriors. We have checked that decreasing the smoothing scale by as much as a factor of 10 has a very marginal impact on the FoMs computed from \cref{fom_definition}. Although the areas of specific 2D posteriors change by $\mathcal{O}(10\%)$, the effect cancels out in the ratios of areas of posteriors which only change by $\mathcal{O}(1\%)$.

\end{itemize}

\section{Results: Constraints from combined probes}\label{sec:combining_probes_constraints}

\subsection{Constraints from the minimal dataset combination and with each extra probe}\label{sec:adding_one_probe}

Constraints on $c_B$ and $c_M$ from our minimal combination of probes are shown in all panels of \crefrange{fig:adding_one_probe}{fig:adding_three_probes}, as these constraints are the baseline against which we compare when we add further probes of DE perturbations. It should be noted that CMB lensing is playing a very significant role in constraining power here. Without CMB lensing, the constraining power of the CMB primary anisotropies alone on late-time perturbations is very weak \cite{Noller:2018wyv}.
\\

{\bf The impact of adding growth rate measurements}:
In the upper left panel of \cref{fig:adding_one_probe}, we show the impact of adding RSD data to our minimal likelihood setup. We find an improvement in the uncertainties on $c_M$ (a 22\% decrease in the posterior width), but no significant change in the constraining power on $c_B$. The overall FoM improves by a factor of 1.41 which is a larger gain than that expected from the product of the 1D marginalised posteriors, due to the fact that lowering $c_M$ weakens one of the two degeneracy directions between $c_B$ and $c_M$ seen in the minimal combination posterior. As explained in \cref{sec:likelihood_combinations}, when including RSD data from DESI DR1, we also swap out the DESI DR2 BAO data in our minimal combination with DESI DR1, to properly account for correlations between RSD and BAO. However, compared to GR, within our EFTofDE model the effect of this choice on the background constraints is very small, in the form of a small increase in uncertainties, as seen from \cref{tab:eft_constraints}. This is due to the fact that in the EFTofDE, deviations of $\{w_0, w_a\}$ away from $\lcdm$ are also constrained by a theoretical prior in the form of the gradient stability condition and the constraining power on perturbations, rather than just the constraining power on the background (see \cref{sec:background_perts} for a detailed investigation of this phenomenon). The impact of RSD data on the 95\% credible region is qualitatively similar to that on the 68\% region.

Growth rate predictions of various EFTofDE models have been compared with data for for nearly a decade \cite{Tsujikawa:2015mga,Perenon:2015sla,Bellini:2015xja,Hu:2016zrh,Kreisch:2017uet,Noller:2018wyv,Perenon:2019dpc,Seraille:2024beb,Shah:2025vnt} It is interesting to see how constraints on the EFTofDE have evolved with the evolution in RSD datasets from earlier compilations including BOSS and several other surveys, to eBOSS, to DESI. We can compare our results quantitatively with works that constrain the $\alpha$ basis parameters with MCMC analyses. In particular, we compare the constraints on $c_M$, since the top left panel of \cref{fig:adding_one_probe} demonstrates that $c_M$ is the parameter for which the posterior is impacted the most by adding RSD data. An early such analysis was performed by \cite{Bellini:2015xja}, but probing a different model space than we do here -- also varying the speed of gravitational waves which strongly prefers to be subluminal for their dataset combination -- as well as considering a large set of RSD measurements also different from the more recent data sets explored here.  
\cite{Noller:2018wyv} then probed the same model space as considered here and used a small set of uncorrelated RSD measurements: specifically from 6dF \cite{Beutler:2012px} and BOSS DR11 CMASS \cite{BOSS:2013yzh}.\footnote{In this analysis, potentially highly correlated BAO measurements -- specifically the BOSS DR11 CMASS anisotropic BAO measurement -- were excluded in order to ensure robustness.}
The resulting constraints on $c_M$ yield $\Delta c_M \sim 1.5$, which is similar to the constraints obtained with the newer eBOSS DR16 RSD and BAO measurements \cite{Howlett:2014opa,BOSS:2016wmc,eBOSS:2020yzd, BOSS:2016wmc, eBOSS:2020hur, eBOSS:2020uxp, eBOSS:2020tmo} in \cite{Shah:2025vnt}, and the ones obtained here with the state-of-the-art DESI DR1 measurements. The high constraining power of early RSD datasets on $c_M$ does not indicate the precision of those datasets but rather their effective redshifts. Since the $\Omega_{DE}$ parametrisation has much stronger phenomenological effects at low redshifts, a growth rate measurement at low redshift is much more constraining on this parametrisation than a measurement of comparable precision at high redshift. In this case, the 6dF RSD measurement is very powerful to constrain $c_M$. In this work, we remain conservative in our RSD dataset choice in the sense that we employ the most recent compilation of RSD and BAO measurements from a single survey, which is DESI DR1 at the time of writing. A much more recent low-redshift measurement of $f\sigma_8$ comes from the DESI peculiar velocity survey \cite{Qin:2025ggt,Lai:2025xkf,Qin:2025rwz}. We leave the use of these measurements to obtain EFTofDE constraints to future work.
\\

{\bf The impact of adding $\bm{\txtp}$ correlations}:
In the upper right panel of \cref{fig:adding_one_probe} we show the impact of replacing the CMB lensing likelihood in our minimal set of probes with DES Y3 $\txtp$ measurements. As explained in \cref{sec:likelihood_combinations}, we remove the CMB lensing likelihood because we expect non-negligible correlations between the CMB lensing and $\txtp$ observables, for which we don't have a model. We see a 27\% reduction in the posterior width of $c_B$ and a reduction of 20\% in the posterior width of $c_M$, with an overall improvement by a factor of 1.71 in the FoM. In the 95\% credible region, the posterior tail on the lower right is reduced but still visible, while it is largely removed in the 68\% region.

The same panel also acts as a comparison of the constraining power of CMB lensing and $\txtp$ on DE perturbations. Evidently, $\txtp$ is a significantly stronger probe than CMB lensing, even when only using data from linear scales. A few factors contribute to this difference: 1) the difference in the amount of information in the two likelihoods, 2) the presence of redshift tomography in the DES $\txtp$ measurements which allows for a higher resolution in redshift as opposed to the CMB lensing likelihood that measures a single redshift integral in the form of the observed lensing convergence of the CMB, and 3) the choice of parametrisation $\alpha_i \propto \Omega_{DE}$, which has weaker effects on phenomenology at higher redshifts, combined with the fact that CMB lensing is sensitive to higher redshifts than $\txtp$.
\\

{\bf The impact of adding the ISW effect}:
In the lower left panel of \cref{fig:adding_one_probe}, we see the effect of adding the CMB $\times$ galaxy surveys ($\ITg$) likelihood to our minimal setup. We find a shift in the means of both $c_B$ and $c_M$ compared to the minimal combination. For $c_B$, the upper and lower uncertainties change, but the change in the posterior width is very marginal (only a 6\% reduction) compared to the minimal combination. The posterior width of $c_M$ is decreased by 24\% compared to the minimal combination. The 2D FoM improves by a factor of 1.54. The $\ITg$ likelihood also significantly affects constraints on the background expansion, and we find that constraints on $w_0-w_a$ are shifted towards $\lcdm$ in all dataset combinations that include $\ITg$, as shown in \cref{tab:eft_constraints}. Apart from the Planck CMB likelihoods, this is the only likelihood in our setup that has a significant impact on the posteriors of both the DE background and perturbations.

In both the 68\% and 95\% regions, we see a reduction in the lower right tail on adding $\ITg$, which is similar to that seen with the addition of the $\txtp$ and $\ITk$ likelihoods. However, a unique feature of adding the $\ITg$ likelihood is an extension of the 95\% credible region in the direction of positively correlated $\cBM$. In fact, in the lower left panel of \cref{fig:adding_one_probe}, the imposed range on the plot (in order to facilitate easier comparison with the other plots) cuts off a very small island of the 95\% credible region that appears on adding $\ITg$ to our minimal combination. We discuss the physical origin of this island and consistency checks in \cref{app-ISW}.

The lower right panel of \cref{fig:adding_one_probe} shows the impact of adding the Planck PR4 ISW ($\ITk$) likelihood to our minimal combination. We find that the posterior mean of $c_B$ is shifted to a lower value, with a 19\% reduction of the posterior width. We also find a shift in the mean $c_M$, and 13\% reduction of the posterior width of $c_M$ compared to the minimal combination. The 2D FoM improves by a factor of 1.42. For the background expansion constraints, we find that adding the $\ITk$ does not have a substantial effect apart from a slight decrease in uncertainties. For the 95\% credible region, the impact of the $\ITk$ likelihood is qualitatively similar to that of the $\txtp$ likelihood as seen from \cref{fig:adding_one_probe}. 
\\

In general, from \cref{fig:adding_one_probe}, it is clear that the constraining power of RSD data is highly complementary to that of $\txtp$ and ISW data. This is due to the fact that RSD data probes clustering of non-relativistic matter which is determined by the Newtonian potential, while $\txtp$ and ISW are stronger probes of the propagation of photons from source galaxies to the CMB, which is determined by the Weyl potential. The Newtonian and Weyl potential are not trivially related in general EFTofDE models, thus providing two phenomenological degrees of freedom. For an understanding of the constraining power of these various probes on parameters that more directly affect the Newtonian and Weyl potentials, we refer the reader to \cref{sec:pheno_constraints}.

\subsection{Multiprobe combinations and complementarity of probes}\label{sec:multiprobe_constraints}

\begin{figure*}[!t]
    \centering
    \begin{subfigure}
        \centering
        \includegraphics[width=0.4\linewidth]{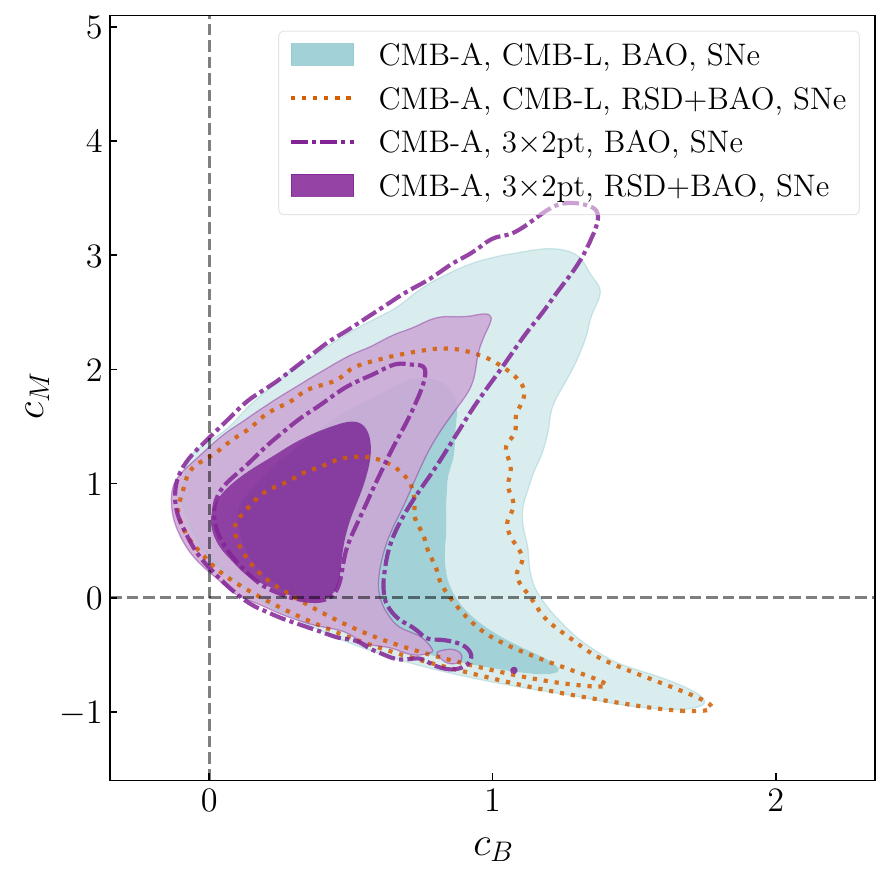}
    \end{subfigure}
    \begin{subfigure}
        \centering
        \includegraphics[width=0.4\linewidth]{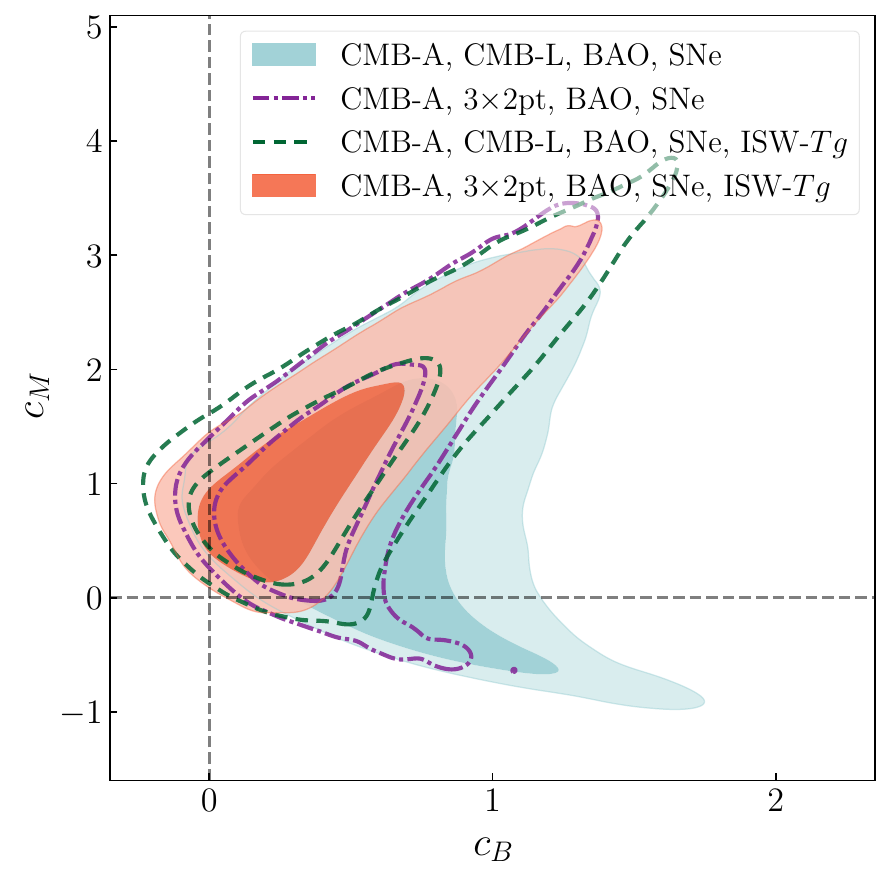}
    \end{subfigure}

    \begin{subfigure}
        \centering
        \includegraphics[width=0.4\linewidth]{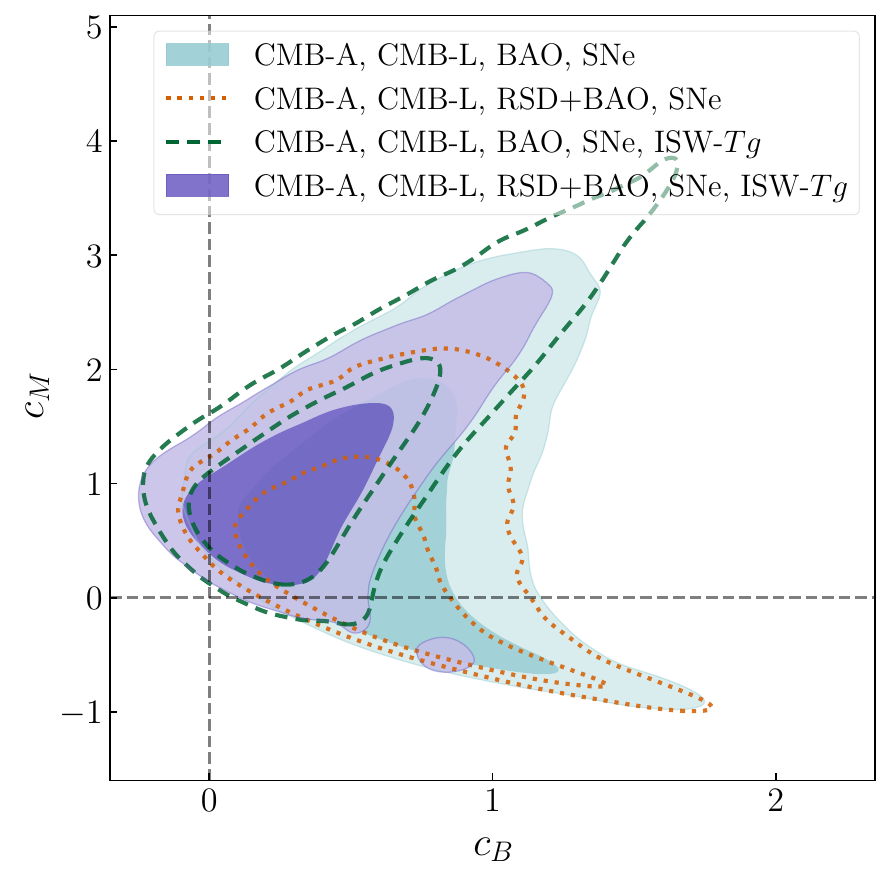}
    \end{subfigure}
    \begin{subfigure}
        \centering
        \includegraphics[width=0.4\linewidth]{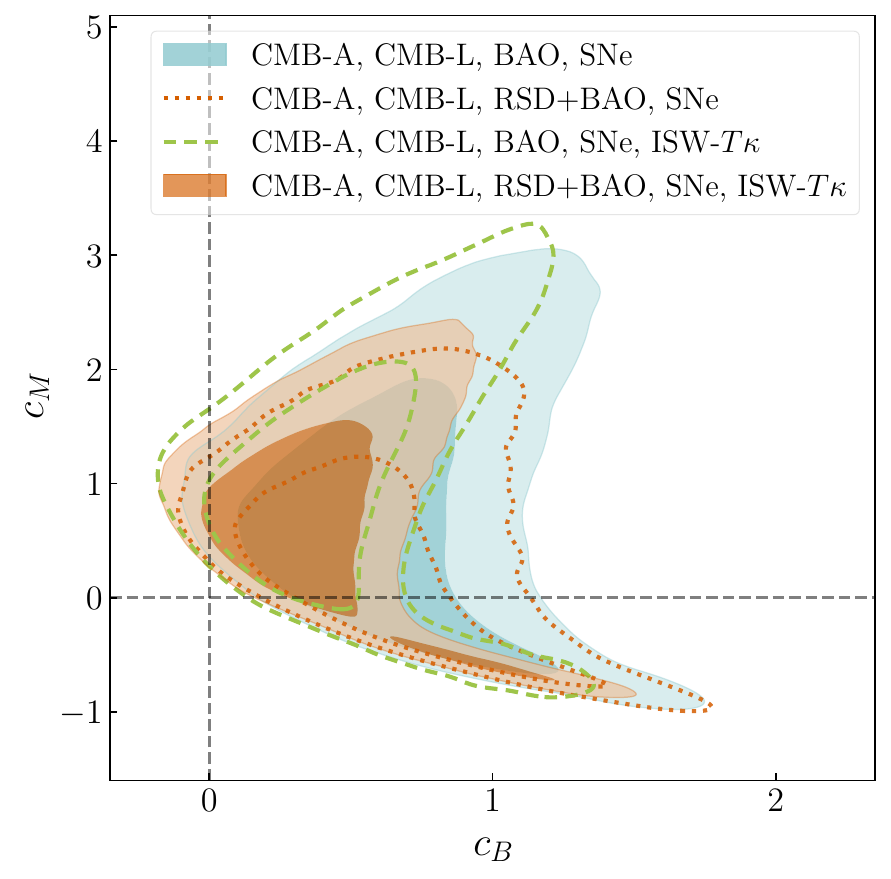}
    \end{subfigure}
    \caption{The impact on the constraints on DE perturbations in the EFTofDE on adding two complementary probes to our minimal combination of probes which comprises of CMB anisotropies (CMB-A) and CMB lensing (CMB-L) from Planck PR4 CamSpec, baryon acoustic oscillations (BAO) from DESI DR2, and Type IA supernovae (SNe) from DES Y5. We overplot the contours from adding each probe individually as shown in \cref{fig:adding_one_probe} to demonstrate the degree of complementarity of the probes. \textbf{Upper left:} Adding DESI DR1 RSD and DES Y3 $\txtp$. \textbf{Upper right:} Adding DES Y3 $\txtp$ and the CMB $\times$ galaxies ISW likelihood ($\ITg$). \textbf{Lower left:} Adding RSD and $\ITg$. \textbf{Lower right:} Adding RSD and the Planck PR4 ISW likelihood ($\ITk$). Note that when adding DESI DR1 RSD, we replace DESI DR2 BAO in our minimal combination with DR1 BAO in order to incorporate the publicly available covariance between DESI DR1 RSD and BAO measurements. Note that when adding DES Y3 $\txtp$, we remove CMB lensing from our minimal combination as the covariance between $\txtp$ and CMB lensing is non-negligible and unmodelled. Also, note that $\ITk$ cannot be directly combined with $\ITg$ or $\txtp$ due to non-negligible covariances (see \cref{sec:likelihood_combinations}).}
    \label{fig:adding_two_probes}
\end{figure*}

In this subsection, we present constraints from combining multiple probes of late-time cosmological perturbations with our probes of the background expansion and CMB lensing. We demonstrate that constraints from different LSS probes are highly complementary, and thus their combination provides stronger and more robust constraints on the entire theory space rather than a single parameter.

The upper left panel of \cref{fig:adding_two_probes} shows the highly complementary nature of the constraining power of RSD and $\txtp$ measurements on DE perturbations. This is a result of the existence of two phenomenological degrees of freedom in the EFTofDE: modifications to structure growth and modifications to lensing. The two degrees of freedom can be represented by the $\mu(z)$ and $\Sigma(z)$ functions in modified Poisson equations as in \cref{Poisson_mu,Poisson_Sigma}. Although $\mu(z)$ and $\Sigma(z)$ are more correlated in the EFTofDE than in purely phenomenological parametrisations \cite{Perenon:2016blf,Peirone:2017ywi,Espejo:2018hxa,Shah:2025vnt}, they are still non-degenerate and thus probes of structure growth and lensing are complementary. Overall, we see a 39\% reduction in the posterior width of $c_B$ and 40\% for $c_M$, with a 2.42 factor improvement in the 2D FoM, when combining RSD and $\txtp$ measurements.

The upper right panel of \cref{fig:adding_two_probes} shows the constraints obtained from combining $\txtp$ measurements with $\ITg$. We see a 26\% reduction in the posterior width of $c_B$ and 34\% for $c_M$, with a 2.24 factor improvement in the 2D FoM, slightly more than the increase anticipated from the simple product of the 1D posterior widths of $c_B$ and $c_M$. This can be understood by there being only one prominent degeneracy direction (a positive degeneracy between $c_B$ and $c_M$) in the posterior compared to two prominent degeneracy directions (the positive degeneracy and a negative degeneracy in the $c_M < 0$ region) in the posterior of the minimal combination. Looking at the 95\% credible regions, the combination of $\txtp$ and $\ITg$ removes the narrow tail at the bottom right of the $\cBM$ posterior which is present in constraints with the minimal combination as well as in constraints with either one of the $\txtp$ or $\ITg$ likelihoods. The $\txtp$ and $\ITg$ likelihoods are both strong probes of lensing but weak probes of clustering, while RSD is a strong clustering probe and weak lensing probe. Thus, the $\txtp$ + $\ITg$ combination affects the $\cBM$ posterior in a different direction than the RSD likelihood.

The lower left panel of \cref{fig:adding_two_probes} shows the constraints obtained from combining RSD measurements with $\ITg$. We see a 22\% reduction in the posterior width of $c_B$ and 38\% for $c_M$, with a 1.95 factor improvement in the 2D FoM. Interestingly, neither of the $\ITg$ or RSD likelihoods improves the $c_B$ posterior width when added individually to the minimal combination, but on combining both of them we do find an improvement in the posterior width of $c_B$.

The lower right panel of \cref{fig:adding_two_probes} shows the constraints obtained from combining RSD measurements with $\ITk$. We see a 27\% reduction in the posterior width of $c_B$ and 20\% for $c_M$, with a 1.86 factor improvement in the 2D FoM, slightly more than the increase anticipated from the simple product of the 1D posteriors of $c_B$ and $c_M$. Note also that despite the 68\% credible region of the smoothed 2D posterior being bimodal, we have found that the two peaks are connected in the unsmoothed posterior.

\begin{figure*}[!t]
    \centering
    \begin{subfigure}
    \centering
    \includegraphics[width=0.4375\linewidth]{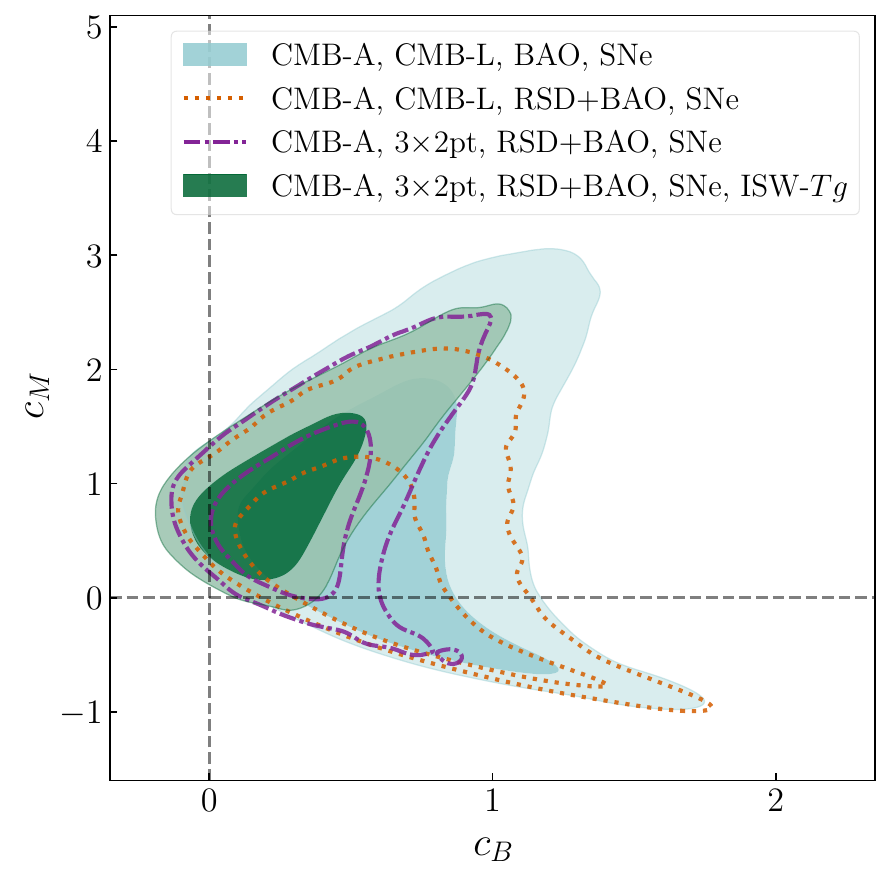}
    \end{subfigure}
    \begin{subfigure}
    \centering
    \includegraphics[width=0.45\linewidth]{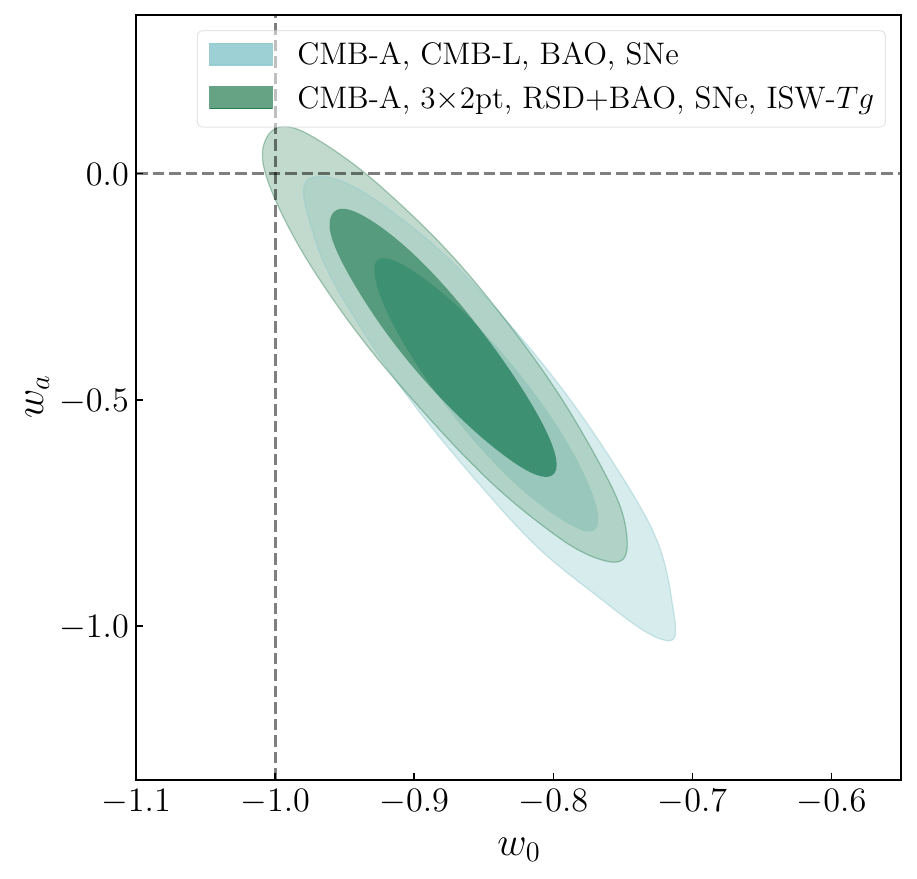}
    \end{subfigure}
    \caption{Strong constraints on the EFTofDE perturbations and background expansion obtained by adding three complementary probes of cosmological perturbations: DESI DR1 RSD, DES Y3 $\txtp$, and the $\ITg$ likelihood \cite{Stolzner:2017ged,Seraille:2024beb} probing the ISW effect. We avoid unmodelled non-negligible covariances between different probes as explained in \cref{sec:likelihood_combinations}. \textbf{Left panel:} Constraints on the parameters controlling linear perturbations $\cBM$. The plot showcases the complementarity of the various probes of structure in the late universe as it shows how they improve constraining power in different directions in the parameter space: adding RSD improves constraints on $c_M$, adding $\txtp$ improves constraints on $c_B$, and adding $\ITg$ eliminates the lower tail of the $\cBM$ posterior, removing a degeneracy direction. \textbf{Right panel:} Constraints on $\wzwa$. The constraints found from the minimal combination of probes comprising of CMB+BAO+SNe incorporate EFTofDE physics and are different from the constraints on $\wzwa$ obtained in the standard $w_0w_a$CDM model from the same dataset combination -- this effect is explained in detail in \cref{sec:background_perts}. The shift in the constraints from the minimal combination to the final probe combination is entirely caused by the constraining power of the $\ITg$ likelihood on the background expansion. \cref{tab:eft_constraints} lists the 1D marginalised constraints on these parameters from all the dataset combinations considered in this work.}
    \label{fig:adding_three_probes}
\end{figure*}

Finally, in \cref{fig:adding_three_probes} we show the constraints obtained from adding RSD, $\txtp$, and $\ITg$ measurements to the minimal combination. This is our most constraining combination of probes of DE perturbations  We see a 35\% reduction in the posterior width of $c_B$ and 44\% for $c_M$, with a 2.69 factor improvement in the 2D FoM. The specific strengths of the three complementary probes are as follows: RSD data increases constraining power on $c_M$ by lowering the upper limit of the posterior, $\txtp$ data cuts off the posterior region in the direction of low $c_M$ and high $c_B$, and $\ITg$ data lowers the mean of the $c_B$ posterior and cuts off the lower right tail of the posterior along low $c_M$ and high $c_B$. The impact of the three probes on the 95\% credible region is qualitatively similar to that on the 68\% region. We also show the corresponding constraints on the expansion history, where the shift in the posterior compared to the minimal combination is primarily caused by the $\ITg$ likelihood, as seen from the $\wzwa$ constraints reported in \cref{tab:eft_constraints}.

\begin{table*}[]
    \centering
\renewcommand{\arraystretch}{1.5}
\begin{tabular}{|c||c|c|c|c|c|}
\hline
\backslashbox{Datasets}{Parameter} & $c_B$ & $c_M$ & $w_0$ & $w_a$ & FoM($\cBM$) \\
\hline\hline
Minimal: CMB-A + CMB-L + BAO + SNe & $0.66^{+0.25}_{-0.37}$ & $0.74^{+0.63}_{-1.1}$ & $-0.847\pm 0.053$ & $-0.50\pm 0.20$ & 1.00 \\
\hline
Minimal + RSD & $0.62^{+0.23}_{-0.36}$ & $0.42^{+0.52}_{-0.82}$ & $-0.847\pm 0.059$ & $-0.53^{+0.25}_{-0.22}$ & 1.41 \\
\hline
Minimal + $\txtp$ & $0.45^{+0.14}_{-0.32}$ & $1.11^{+0.49}_{-0.92}$ & $-0.846\pm 0.052$ & $-0.49\pm 0.19$ & 1.71 \\
\hline
Minimal + $\ITg$ & $0.49^{+0.13}_{-0.45}$ & $1.37^{+0.29}_{-1.0}$ & $-0.872\pm 0.053$ & $-0.39^{+0.18}_{-0.21}$ & 1.54 \\
\hline
Minimal + $\ITk$ & $0.44^{+0.17}_{-0.33}$ & $1.04^{+0.64}_{-0.86}$ & $-0.850\pm 0.051$ & $-0.48\pm 0.18$ & 1.42 \\
\hhline{|=||=|=|=|=|=|}
Minimal + RSD + $\txtp$ & $0.35^{+0.14}_{-0.24}$ & $0.82^{+0.40}_{-0.64}$ & $-0.850\pm 0.056$ & $-0.50\pm 0.21$ & 2.42 \\
\hline
Minimal + $\ITg$ + $\txtp$ & $0.41^{+0.09}_{-0.37}$ & $1.24^{+0.25}_{-0.90}$ & $-0.870\pm 0.052$ & $-0.39^{+0.18}_{-0.20}$ & 2.24 \\
\hline
Minimal + RSD + $\ITg$ & $0.35^{+0.16}_{-0.32}$ & $1.01^{+0.38}_{-0.70}$ & $-0.882\pm 0.054$ & $-0.37\pm 0.20$ & 1.95 \\
\hline
Minimal + RSD + $\ITk$ & $0.40^{+0.15}_{-0.30}$ & $0.74\pm 0.69$ & $-0.856\pm 0.055$ & $-0.49\pm 0.21$ & 1.86 \\
\hhline{|=||=|=|=|=|=|}
Minimal + RSD + $\ITg$ + $\txtp$& $0.29^{+0.13}_{-0.27}$ & $0.95^{+0.34}_{-0.63}$ & $-0.880\pm 0.053$ & $-0.37\pm 0.19$ & 2.69 \\
\hhline{|=||=|=|=|=|=|}
Minimal, $\lcdm$ background & $0.96^{+0.30}_{-0.43}$ & $1.7^{+1.0}_{-1.6}$ & - & - & 0.69 \\
\hline
CMB-A, CMB-L, $\lcdm$ background & $0.74^{+0.30}_{-0.43}$ & $1.29^{+0.73}_{-1.4}$ & - & - & 0.74 \\
\hline
Minimal + RSD + $\ITg$, $\lcdm$ background & $0.47^{+0.27}_{-0.31}$ & $1.42^{+0.65}_{-0.76}$ & - & - & 1.62 \\
\hline
\hhline{|=||=|=|=|=|=|}
Minimal (GR) & - & - & $-0.807\pm 0.056$ & $-0.73^{+0.23}_{-0.21}$ & - \\
\hline
CMB-early + BAO + SNe & - & - & $-0.818\pm 0.057$ & $-0.64^{+0.24}_{-0.22}$ & - \\
\hline
\end{tabular}

    \caption{Constraints on the parameters describing modified gravity perturbations $\cBM$ and background expansion $\wzwa$ for all the dataset combinations discussed in the main text. This table shows the means and the 68\% uncertainties on the parameters. Our FoM for $\cBM$ is defined as the area of the 68\% credible region of their 2D posterior -- see \cref{sec:quantifying_constraining_power} for more details. The abbreviations for the datasets here refer to the following -- CMB-A: Planck PR4 CMB anisotropies, CMB-L: Planck PR4 CMB lensing, BAO: DESI DR2, SNe: DES-Dovekie, RSD+BAO: DESI DR1 joint RSD+BAO, $\txtp$: DES Y3, $\ITg$: ISW from CMB-galaxy cross-correlations, $\ITk$: ISW from CMB temperature and CMB lensing cross-correlations. See \cref{sec:data} for details on the datasets.}
    \label{tab:eft_constraints}
\end{table*}

\subsection{Constraints on modified gravity phenomenology}\label{sec:pheno_constraints}

Although the functional degrees of freedom in the EFTofDE are $\{\alpha_B(z), \alpha_M(z)\}$, the combinations of these parameters that affect the phenomenology more directly are the ones that appear in the Poisson equations \cref{Poisson_mu,Poisson_Sigma}: namely $\{\mu(z), \Sigma(z)\}$. As seen from \cref{mu_qsa,Sigma_qsa}, they have a highly nontrivial dependence on $\{\alpha_B(z), \alpha_M(z)\}$ and are also sensitive to their time derivatives via $c_s^2$ and past evolution via $M^2$. This translates to a highly nonlinear dependence on the varied parameters $\cBM$. We describe the constraints obtained on modified gravity phenomenology in the EFTofDE model with our various probes in two ways: for a quantitative description, we compute the posterior widths and FoMs for the present day values of $\{\mu(z), \Sigma(z)\}$, and for a more qualitative yet physically insightful description, we show how each of our probes of clustering and lensing disfavours a particular regime of modified gravity phenomenology by disfavouring specific time evolutions of $\{\mu(z), \Sigma(z)\}$.

\begin{figure}
    \centering
    \includegraphics[width=\linewidth]{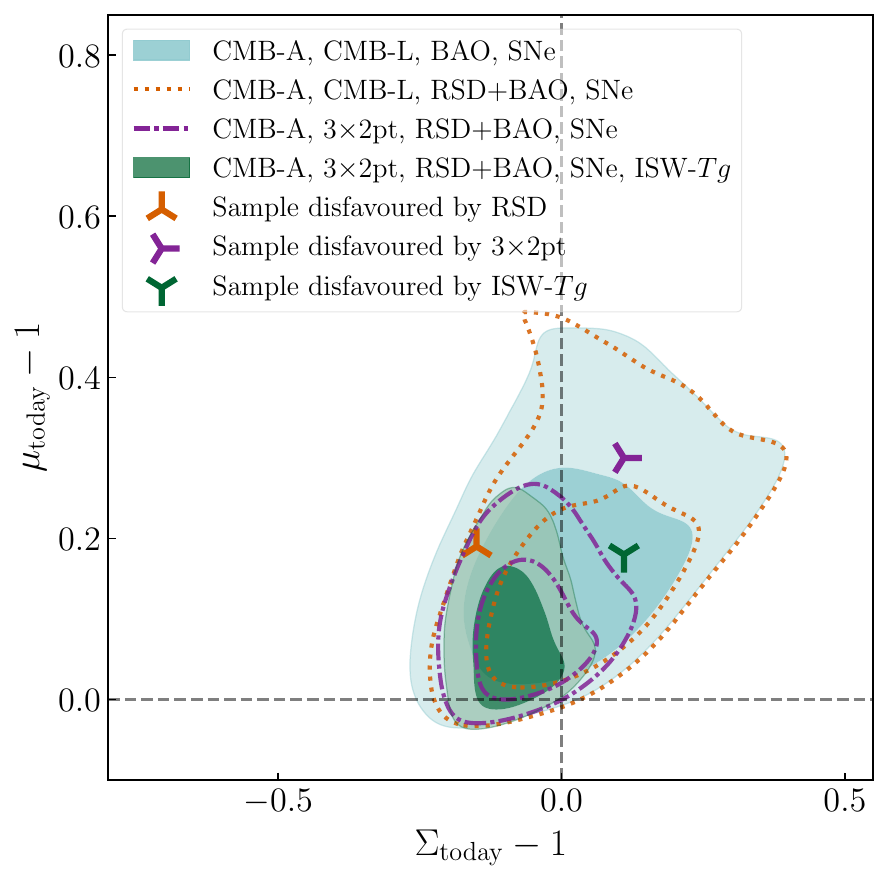}
    \caption{Constraints on the present-day values $\mSt$ of the phenomenological parameters describing modifications to the evolution of gravitational potentials in the EFTofDE according to \cref{Poisson_mu,Poisson_Sigma}. As shown earlier in \cref{fig:adding_three_probes}, we show how each individual probe contributes to the constraints from the minimal combination of probes, resulting in strong constraints obtained with the combination with the most probes. We also show how an individual sample disfavoured by each extra probe individually in the $\cBM$ plane, shown in \cref{fig:adding_one_probe}, maps to the phenomenological $\mSt$ plane. Constraints from all the various dataset combinations considered in this work are in \cref{tab:musigma}. These results are elaborated and complemented by looking at the past evolution of $\{\mu(z), \Sigma(z)\}$ in \cref{sec:pheno_constraints} and \cref{fig:newprobes_disfavoured_musigma_evolution}.}
    \label{fig:minimal_rsd_isw_3x2pt_musigma}
\end{figure}

We consider the present day values of $\{\mu(z), \Sigma(z)\}$ denoted $\mSt$ (analogous to how $\cBM$ are related to the present day values of $\{\alpha_B(z), \alpha_M(z)\}$ but without an extra $\Omega_{DE}$ factor), and we compute the 1D marginalised constraints and FoM values for the phenomenological parameters. In \cref{fig:minimal_rsd_isw_3x2pt_musigma}, we show the significant improvement in the constraints on $\mSt$ obtained by adding RSD, $\txtp$, and $\ITg$ measurements to our minimal combination. The combination of various probes of cosmic structure places the observational constraints firmly in the regime of $\mt > 1,\ \St < 1$, as was also noted in \cite{Shah:2025vnt}, and is interesting considering that in the prior space of $\cBM$ it is far more likely to have $\mt < 1,\ \St < 1$ or $\mt > 1,\ \St > 1$ \cite{Pogosian:2016ji}. There is an observational precedent for probes of lensing indicating a suppression in either structure growth or gravity: this is captured by the so-called $S_8$ tension within $\Lambda$CDM cosmology, referring to lower values of the parameter $S_8 \equiv \sigma_8(\Omega_m/0.3)^{0.5}$ (see \cite{CosmoVerseNetwork:2025alb} for a recent review) and by direct measurements of the Weyl potential inferring values below the predictions of GR \cite{Tutusaus:2023aux,Rosatello:2026znt}. The FoM for $\mSt$ computed using the area of the 68\% contour (the same as we do for $\cBM$) for the combination of RSD, $\txtp$ and $\ITg$ is higher by a factor of 3.37 compared to the FoM of the minimal combination. \cref{tab:musigma} shows the 1D marginalised constraints and FoMs of $\mSt$ for all our data combinations. In general, we see that the improvements in FoMs of $\mSt$ in \cref{tab:musigma} for most data combinations are higher than the improvements on $\cBM$ for the same data combinations shown in \cref{tab:eft_constraints}. 

\begin{figure*}[!t]
    \centering
    \begin{subfigure}
        \centering
        \includegraphics[width=0.45\linewidth]{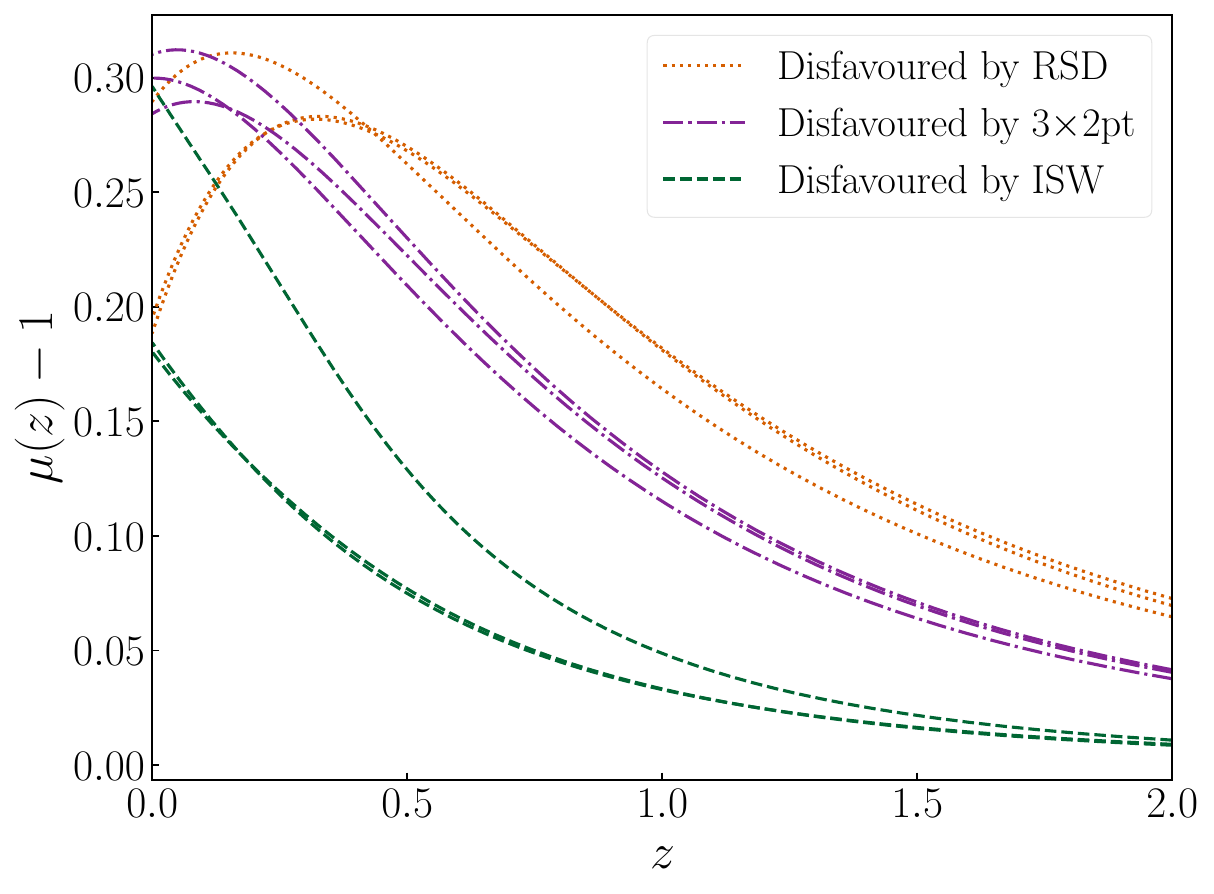}    
    \end{subfigure}
    \begin{subfigure}
        \centering
        \includegraphics[width=0.4615\linewidth]{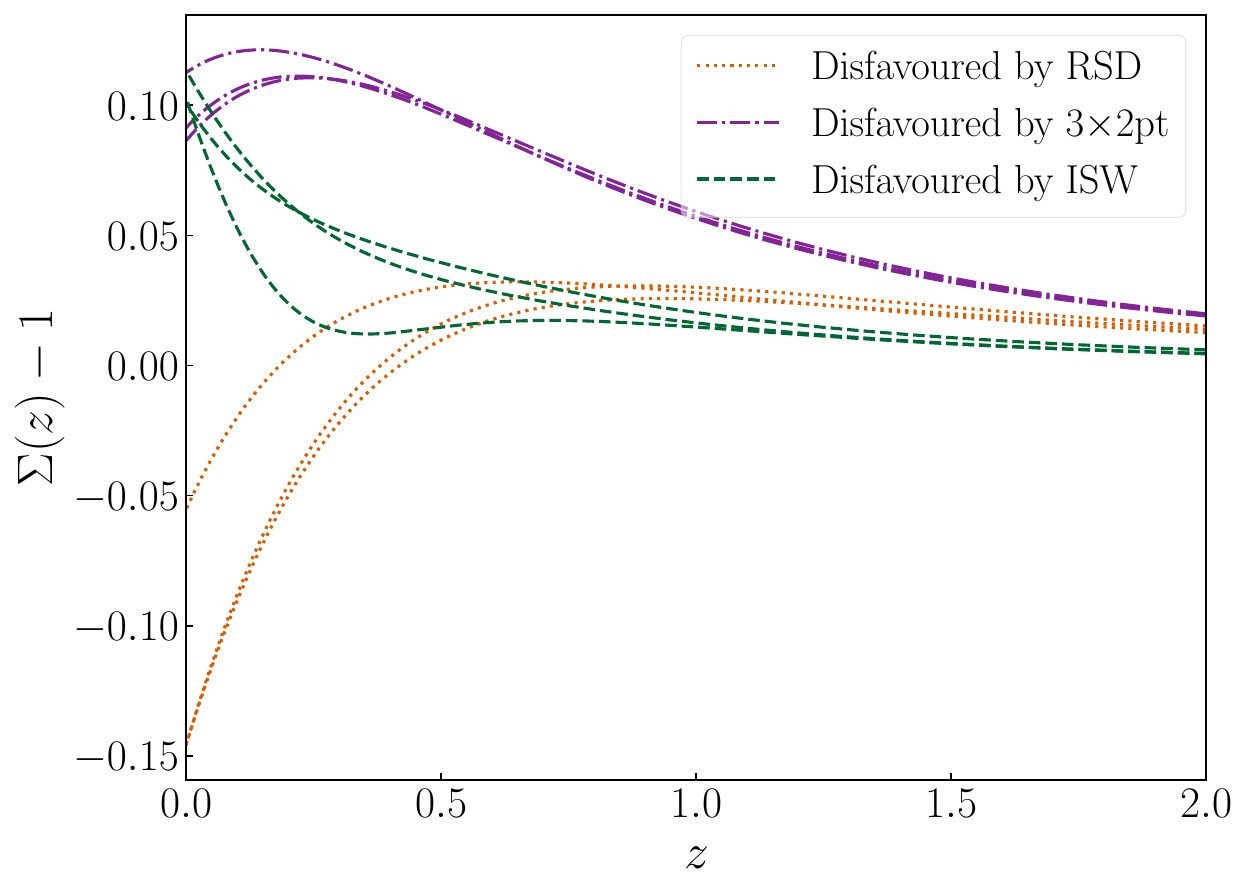}
    \end{subfigure}
    \caption{The cosmic evolutions of $\{\mu(z), \Sigma(z)\}$ for some samples within the 95\% posteriors in the minimal combination that are disfavoured by each extra probe -- these are the samples shown in the $\cBM$ plane in \cref{fig:adding_one_probe} along with a few samples with nearby $\cBM$ values. The major conclusions from these plots are: 1) RSD data disfavours strong enhancements to matter clustering over cosmic history as shown by the left panel, 2) $\txtp$ data disfavours strong enhancements to lensing over cosmic history as shown by the right panel, 3) $\ITg$ data constraints the sign of the time evolution of the Weyl potential $\Phi+\Psi$ at low redshifts $z\ \lsim\ 0.5$ as shown by the right panel. Also, different points in the $\cBM$ plane can map to very similar $\mSt$ values as shown by both panels.}
    \label{fig:newprobes_disfavoured_musigma_evolution}
\end{figure*}

The present-day values of $\mSt$ are not always a good proxy for their evolution across cosmic history since in EFTofDE models $\{\mu(z), \Sigma(z)\}$ may not have a monotonic evolution and may even cross unity, switching between enhancement and suppression of gravity. It is thus useful to look at the evolution of $\{\mu(z), \Sigma(z)\}$ across redshifts. In particular, we can see how each of our cosmic probes disfavours a particular regime of modified gravity phenomenology. In \cref{fig:adding_one_probe}, we indicate the particular points in the $\cBM$ parameter space that are allowed in our minimal constraints but disfavoured by each of the RSD, $\txtp$ and $\ITg$ likelihoods. We identify the particular features in the time evolution of $\mu(z)$ or $\Sigma(z)$ corresponding to these parts of the parameter space, and plot these in \cref{fig:newprobes_disfavoured_musigma_evolution} to show how each individual probe of cosmic structure constrains modified gravity phenomenology. In particular, we find that the RSD data rules out $\mu(z)\ \gsim\ 0.2$ for an extended period in the redshift range $0.3\ \lsim\ z\ \lsim\ 1.5$ (the range of effective redshifts of the DESI DR1 RSD likelihood \cite{DESI:2024jxi}), the $\txtp$ data rules out $\Sigma(z)\ \gsim\ 0.1$ at low redshifts $z\ \lsim\ 2$, and the $\ITg$ data rules out negative values of $\frac{d}{dz}\Sigma(z)$ (ie. higher $\Sigma(z)$ at later times) at very low redshifts $z\ \lsim\ 0.5$.

\begin{table*}[]
    \centering
\renewcommand{\arraystretch}{1.5}
\begin{tabular}{|c||c|c|c|}
\hline
\multirow{2}{*}{Datasets} & $\mt$ & $\St$ & FoM \\
& & & ($\mt, \St$) \\
\hline\hline
Minimal: CMB-A + CMB-L + BAO + SNe & $0.18^{+{0.06}}_{-{0.13}}$ & $0.02^{+{0.12}}_{-{0.13}}$ & 1.00 \\
\hline
Minimal + RSD & $0.17^{+{0.05}}_{-{0.12}}$ & $0.05\pm 0.11$ & 1.29 \\
\hline
Minimal + ISW-$Tg$ & $0.12^{+{0.04}}_{-{0.12}}$ & $-0.09^{+{0.06}}_{-{0.07}}$ & 1.94 \\
\hline
Minimal + ISW-$T\kappa$ &$0.11^{+{0.04}}_{-{0.09}}$ & $-0.08^{+{0.08}}_{-{0.10}}$ & 1.70 \\
\hline
Minimal + 3$\times$2pt & $0.11^{+{0.04}}_{-{0.09}}$ & $-0.07^{+{0.07}}_{-{0.08}}$ & 2.09 \\
\hhline{|=||=|=|=|}
Minimal + RSD + ISW-$Tg$ & $0.10^{+{0.04}}_{-{0.09}}$ & $-0.08^{+{0.06}}_{-{0.07}}$ & 2.34 \\
\hline
Minimal + RSD + ISW-$T\kappa$ & $0.11^{+{0.03}}_{-{0.10}}$ & $-0.05^{+{0.08}}_{-{0.09}}$ & 2.03 \\
\hline
Minimal + RSD + 3$\times$2pt & $0.09^{+{0.04}}_{-{0.08}}$ & $-0.06^{+{0.06}}_{-{0.07}}$ & 2.70 \\
\hline
Minimal + ISW-$Tg$ + 3$\times$2pt & $0.10^{+{0.04}}_{-{0.10}}$ & $-0.09^{+{0.05}}_{-{0.05}}$ & 2.84 \\
\hline
\hhline{|=||=|=|=|}
Minimal + RSD + ISW-$Tg$ + 3$\times$2pt & $0.08^{+{0.04}}_{-{0.08}}$ & $-0.09^{+{0.05}}_{-{0.05}}$ & 3.37 \\
\hline
\hhline{|=||=|=|=|}
CMB-early + BAO + SNe & - & - & - \\
\hline
Minimal, $\lcdm$ background & $0.29^{+{0.09}}_{-{0.12}}$ & $0.01^{+{0.12}}_{-{0.16}}$ & 0.74 \\
\hline
Minimal + RSD + ISW-$Tg$, $\lcdm$ background & $0.17\pm0.07$ & $-0.08^{+{0.06}}_{-{0.07}}$ & 2.24 \\
\hline
\end{tabular}
    \caption{Constraints on the phenomenological parameters $\mSt$ for all the dataset combinations discussed in the main text. This table shows the means and the 68\% uncertainties on the parameters. Our FoM for $\mSt$ is defined as the area of the 68\% credible region of their 2D posterior -- see \cref{sec:quantifying_constraining_power} for more details. The abbreviations for the datasets here refer to the following -- CMB-A: Planck PR4 CMB anisotropies, CMB-L: Planck PR4 CMB lensing, BAO: DESI DR2, SNe: DES-Dovekie, RSD+BAO: DESI DR1 joint RSD+BAO, $\txtp$: DES Y3, $\ITg$: ISW from CMB-galaxy cross-correlations, $\ITk$: ISW from CMB temperature and CMB lensing cross-correlations. See \cref{sec:data} for details on the datasets.}
    \label{tab:musigma}
\end{table*}

\section{Results: Interplay of background and perturbation constraints}\label{sec:background_perts}

In the previous section, we have looked at how the effects of modified gravity on cosmological perturbations are constrained by various probes of cosmological perturbations. However, there are several ways in which constraints on DE perturbations are also impacted by assumptions and constraints on the background expansion. In this section, we discuss how constraints on DE perturbations are affected by the choice of the background expansion model, by the choice of datasets constraining the background expansion, and by a theoretical prior on $\cBM$ coming from the physical constraint of gradient stability relating the DE background and perturbations. 

\subsection{Impact of background analysis choices}\label{sec:background_choices}

\begin{figure}[t]
    \centering
    \includegraphics[width=\linewidth]{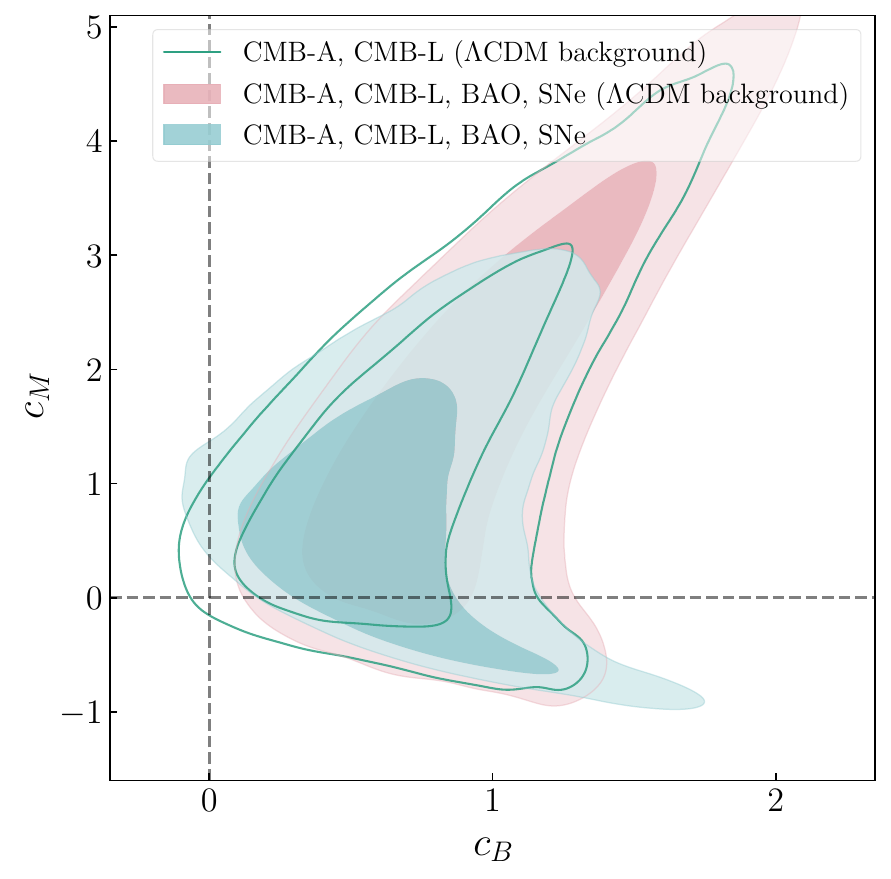}
    \caption{Comparison of constraints on $\cBM$ for the minimal dataset combination across $\lcdm$ and $w_0w_a$ background choices. For the $\lcdm$ background, we also show the impact of the BAO and SNe datasets by comparing constraints with and without including them. We find that assuming a simpler $\lcdm$ expansion history weakens the constraints on $\cBM$ and moves the posterior further away from the GR limit. We also find that removing BAO and SNe datasets makes the $\cBM$ consistent with the GR limit within 95\% when a $\lcdm$ background is assumed.}
    \label{fig:cmbonly_minimal_bm_w0wa_lcdmbackground}
\end{figure}

\begin{figure}[t]
    \centering
    \includegraphics[width=\linewidth]{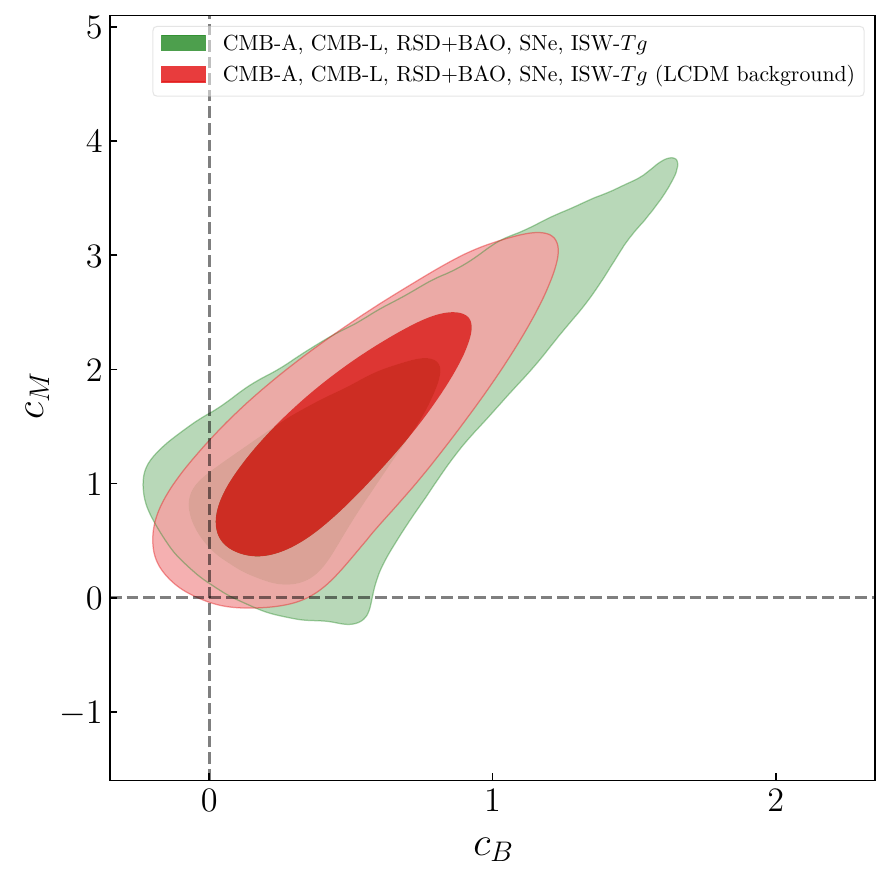}
    \caption{Comparison of constraints on $\cBM$ for the minimal+RSD+$\ITg$ dataset combination across $\lcdm$ and $w_0w_a$ background choices. We find that assuming a simpler $\lcdm$ expansion history changes the shape of the posterior and makes $\cBM$ consistent with the GR limit within 95\%.}
    \label{fig:minimal_rsd_isw_bm_lcdmbackground}
\end{figure}

From \cref{fig:cmbonly_minimal_bm_w0wa_lcdmbackground,fig:minimal_rsd_isw_bm_lcdmbackground}, we see that we get very different posteriors for $\cBM$ based on the assumed form for the background expansion history. This is partly caused by the increased precision in recent expansion history constraints -- this conclusion differs from the conclusion obtained recently in \cite{Shah:2025vnt}, where there was only a small change in the constraints on $\mSt$ when the constraints on the background expansion came from the eBOSS DR16 BAO, Pantheon+ supernovae, and Planck PR3 CMB observations. In our dataset choices for this work, the major difference in expansion history constraints compared to \cite{Shah:2025vnt} comes from the DESI DR2 BAO and DES-Dovekie SNe datasets replacing eBOSS DR16 and Pantheon+ respectively. For the minimal dataset combination in \cref{fig:cmbonly_minimal_bm_w0wa_lcdmbackground}, the surprising feature is that the 68\% credible region on $\cBM$ is tighter with the $w_0w_a$ background, despite this model having two extra parameters marginalised over compared to the $\lcdm$ background. 

Another feature illustrating the sensitivity of the $\cBM$ posterior to expansion history constraints is the impact of removing the BAO and SNe datasets, as shown in \cref{fig:cmbonly_minimal_bm_w0wa_lcdmbackground}. The BAO and SNe datasets are exclusively probes of the background expansion history and have no sensitivity to cosmological perturbations, while $\cBM$ only affect linear perturbations and have no effect on the expansion history. However, \cref{fig:cmbonly_minimal_bm_w0wa_lcdmbackground} shows that these datasets strongly affect the posteriors $\cBM$ even when a $\lcdm$ expansion history is assumed and there is very little phenomenological freedom in the background expansion. We also see that when using the CMB data on its own, the $\cBM$ posterior is consistent with GR at 95\% C.L., but it is not consistent when BAO and SNe data are added. In order to explain this impact of the BAO and SNe data even in a $\lcdm$ background, we turn to the other cosmological parameters. A notable feature of the combination of datasets used here: Planck PR4 CMB, DESI DR2 BAO, and DES-Dovekie SNe, is the $\Omega_m$ discrepancy between Planck CMB and DESI BAO \cite{DESI:2025zgx}. In the context of Parametrized Post-Friedmann (PPF) dynamical DE models, this discrepancy is at the $\sim 2\sigma$ level when a $\lcdm$ background is assumed, but is alleviated in a $w_0w_a$ background. In \cref{app-Omegam} we confirm the presence of the $\Omega_m$ discrepancy in our EFTofDE model with a $\lcdm$ background, and show that there is a noticeable degeneracy between $c_B$ and $\Omega_m$. This explains why the inclusion of BAO and SNe data has a significant effect on the posterior of $c_B$, and shows that the shift in $\Omega_m$ between the dataset combinations leads to a shift in $c_B$ impacting its consistency with the GR limit.

\subsection{The gradient stability prior}\label{sec:stability_prior}

A feature worth noting in all the posteriors shown in \crefrange{fig:adding_one_probe}{fig:adding_three_probes} is that the lower left direction of the posteriors, facing the GR limit, is the direction least affected by the choice of likelihoods. For all the likelihood choices considered in \crefrange{fig:adding_one_probe}{fig:adding_three_probes}, GR remains outside 95\% C.L. In fact, GR lies within the 95\% C.L. for only two analysis choices, both of which involve assuming a $\lcdm$ background: with only the CMB data as shown in \cref{fig:cmbonly_minimal_bm_w0wa_lcdmbackground}, and with the Minimal + RSD + $\ITg$ data combination shown in \cref{fig:minimal_rsd_isw_bm_lcdmbackground}. It appears that in this direction of the parameter space, constraints on DE perturbations are far more sensitive to information about the background expansion than about the perturbations themselves. This indicates that this lower left boundary of the posterior is not determined by the data but by a prior effect coupling the background expansion to the DE perturbations. Since the priors we impose by hand on the dark energy parameters $\{c_B, c_M, w_0, w_a\}$ are wide, uninformative, and uncorrelated, this boundary is explained by a theoretical prior which we describe here in detail.

In the EFTofDE, the linear perturbations of the scalar field are described by a sound speed $c_s$ given by \cref{cs2}. $c_s^2$ depends nonlinearly on $\{c_B, c_M, w_0, w_a\}$ through the time evolutions of $\alpha_B, \alpha_M, M^2,$ and $H$. The sound speed squared needs to remain positive, as a negative value represents exponentially growing perturbations which would be catastrophic for predictions of growth of structure. The requirement of positive sound speed at all times is called the gradient stability condition\footnote{The sound speed squared can in principle be negative for short periods at early times, however this only happens for a very small subregion of the parameter space. Strictly enforcing $c_s^2 > 0$ at all times thus has no significant impact on the posteriors, as checked by e.g. \cite{Noller:2018wyv}.}. In canonical and k-essence scalar field models (which the EFTofDE reduces to in the limit $(c_B, c_M) = (0, 0)$),
the expression \cref{cs2} reduces to

\begin{gather}
 c_s^2(z) = \frac{3(1+w)(1-\Omega_{DE})}{D}
\end{gather}
where redshift dependences are suppressed on the RHS. In this case, a phantom crossing in $(w_0, w_a)$ is not possible as it leads to a violation of either the gradient stability ($c_s^2 \geq 0$) or non-ghost condition ($D > 0$). In this work, we fixed $\alpha_K$ to be positive so that the positivity of $D = \alpha_K + \frac{3}{2}\alpha_B^2$ is ensured. Thus, a phantom crossing leads to the gradient instability. To obtain a stable phantom crossing evolution in the context of the EFTofDE therefore requires going beyond canonical and k-essence scalar field models, so e.g. including some of the higher order (in derivatives) interactions also contained within the Horndeski class of scalar field models.

\begin{figure}[t]
    \centering
    \includegraphics[width=\linewidth]{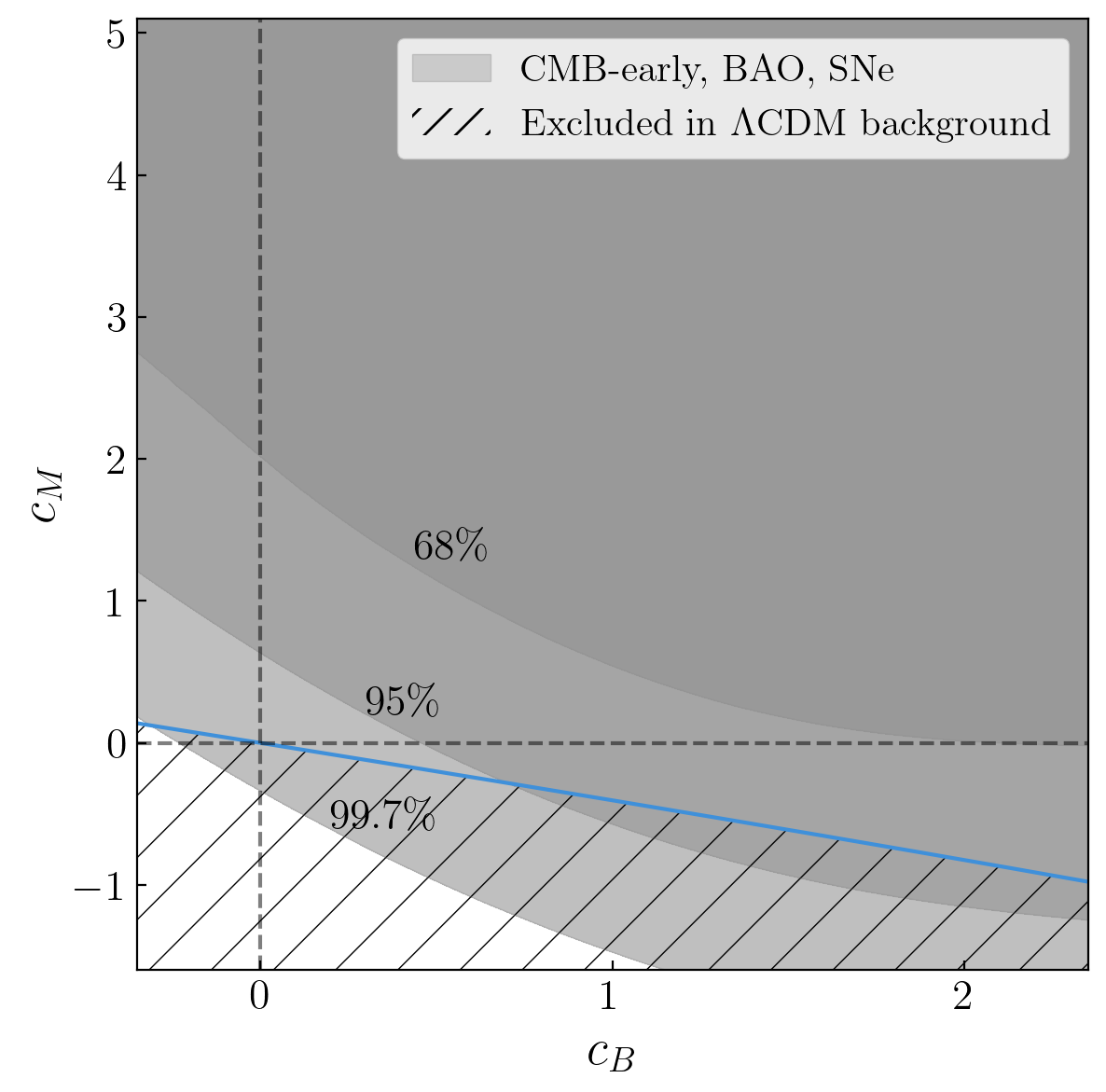}
    \caption{The posterior on $\cBM$ obtained with the dataset combination of the early-universe Planck prior, DESI DR2 BAO, and DES-Dovekie SNe. Despite $\cBM$ having no \togglered{phenomenological} impact on the expansion history and the dataset combination having no sensitivity to cosmological perturbations, we see a nontrivial posterior shape with a mild disfavouring of the GR limit. \togglered{This is due to correlations induced between $\cBM$ and the expansion history via theoretical prior of gradient stability requiring the positivity of the sound speed in the EFTofDE (\cref{cs2}), as explained in detail in \cref{sec:stability_prior}.} The constraints on the expansion history from the same dataset combination are shown in \cref{fig:backgroundonly_w0wa}.}
    \label{fig:backgroundonly_bm}
\end{figure}

For a completely fixed expansion history, such as a $\lcdm$ background, the gradient stability condition is a sharp 1D cutoff in the $\cBM$ plane, as shown in \cref{fig:backgroundonly_bm}. However, when $\wzwa$ are varied parameters that are also constrained by data alongide $\cBM$, the boundary separating the allowed and forbidden parameter spaces of the gradient stability condition is a 3D surface in the 4D region parameter space of $\{c_B, c_M, w_0, w_a\}$\footnote{Note that \cref{cs2} also depends on $\Omega_{m,0}$, which can be considered a dark energy parameter as it is directly related to the background energy density of dark energy as $1 - \Omega_{DE,0}$. However, it is constrained far more tightly than $\{c_B, c_M, w_0, w_a\}$ (despite the $\Omega_m$ discrepancy discussed in the context of \cref{fig:Omega_m_discrepancy}) and its variation within its observed posterior has a negligible impact on \cref{cs2}.}. We are concerned with the effect of this boundary on the observed 2D marginalised posteriors of $\cBM$. To isolate this effect, we can obtain the posterior on $\cBM$ with a likelihood combination similar to our minimal setup but without any constraining power on dark energy perturbations. This gives us the projection in the $\cBM$ plane of the region of the prior boundary that is actually in contact with the observed posteriors of the remaining parameters $\wzwa$. We achieve this by substituting the CMB and CMB lensing likelihoods in our minimal setup with a compressed prior on the early-universe parameters obtained from chains run with the full CMB likelihoods provided in \cite{Lemos:2023xhs}.

\cref{fig:backgroundonly_bm} shows that the observed posteriors for the cosmic expansion history from the CMB, BAO, and SNe data induce a preference for $\cBM$ away from the GR limit without any constraining power on dark energy perturbations. This is a constraint on $\cBM$ in a setup where they don't affect any observables; it is the theoretical prior on these parameters which is implicitly \togglered{present} in our analyses \togglered{due to the requirement of satisfying the gradient stability condition}. \togglered{
When dark energy perturbations are explicitly modelled as a propagating degree of freedom, one generally requires the scalar perturbations to satisfy the no-gradient-instability condition, since otherwise perturbations grow exponentially on sufficiently small scales, leading to a breakdown of the perturbative description \footnote{\togglered{In a numerical context, it then makes sense to speak of "imposing" gradient stability because the parameter space with exponentially growing perturbations is much slower to evaluate in Boltzmann codes, hence it is useful to eliminate it a priori as it is known to be unphysical.}}. Violations may be tolerated only if the resulting instability is sufficiently weak or short-lived that it does not affect the relevant observables.
 \cref{fig:backgroundonly_bm} serves to isolate the impact of this condition on the posteriors of $\cBM$.}

\begin{figure}[t]
    \centering
    \includegraphics[width=\linewidth]{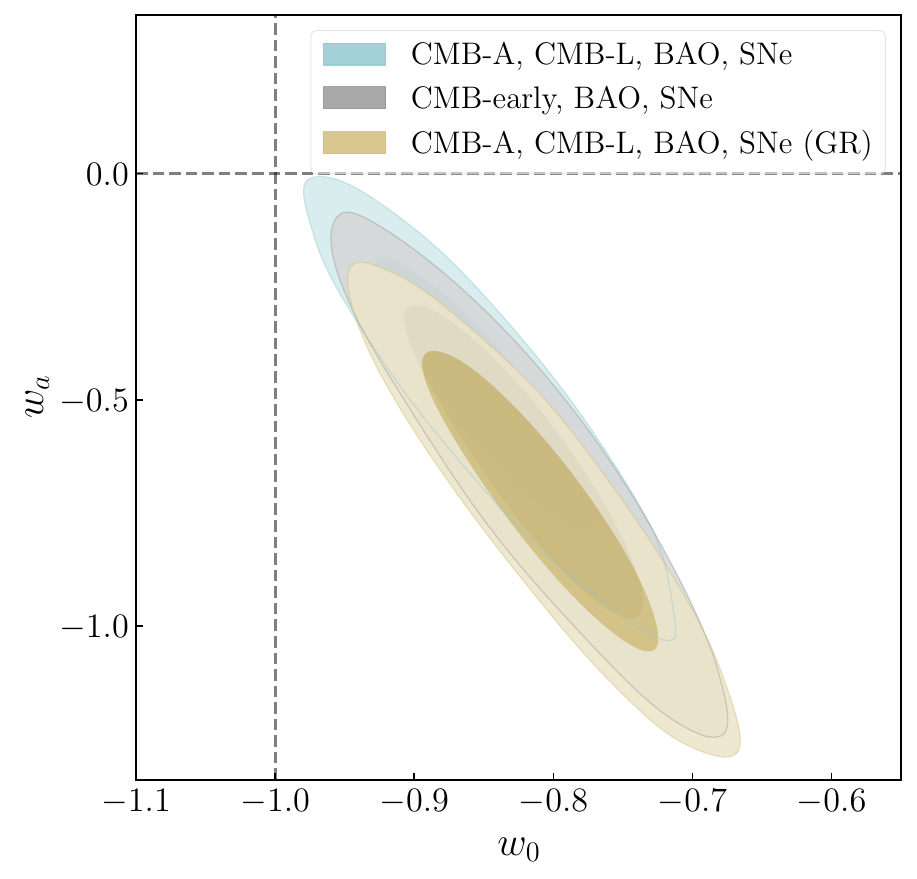}
    \caption{Comparison of the constraints on $\wzwa$ across 1) Our reference case of the minimal dataset combination and the assumption of the EFTofDE model, 2) replacing the full CMB anisotropy and lensing likelihoods in the reference case with the Planck early-universe prior, which removes all sensitivity to linear perturbations in the dataset combination and makes it only sensitive to the background expansion, 3) Changing the model in our reference case from EFTofDE to the standard $w_0w_a$CDM which uses the PPF formalism to allow for phantom crossing and neglects subhorizon DE perturbations.}
    \label{fig:backgroundonly_w0wa}
\end{figure}

\begin{figure*}[t]
    \centering
    \includegraphics[width=0.9\textwidth]{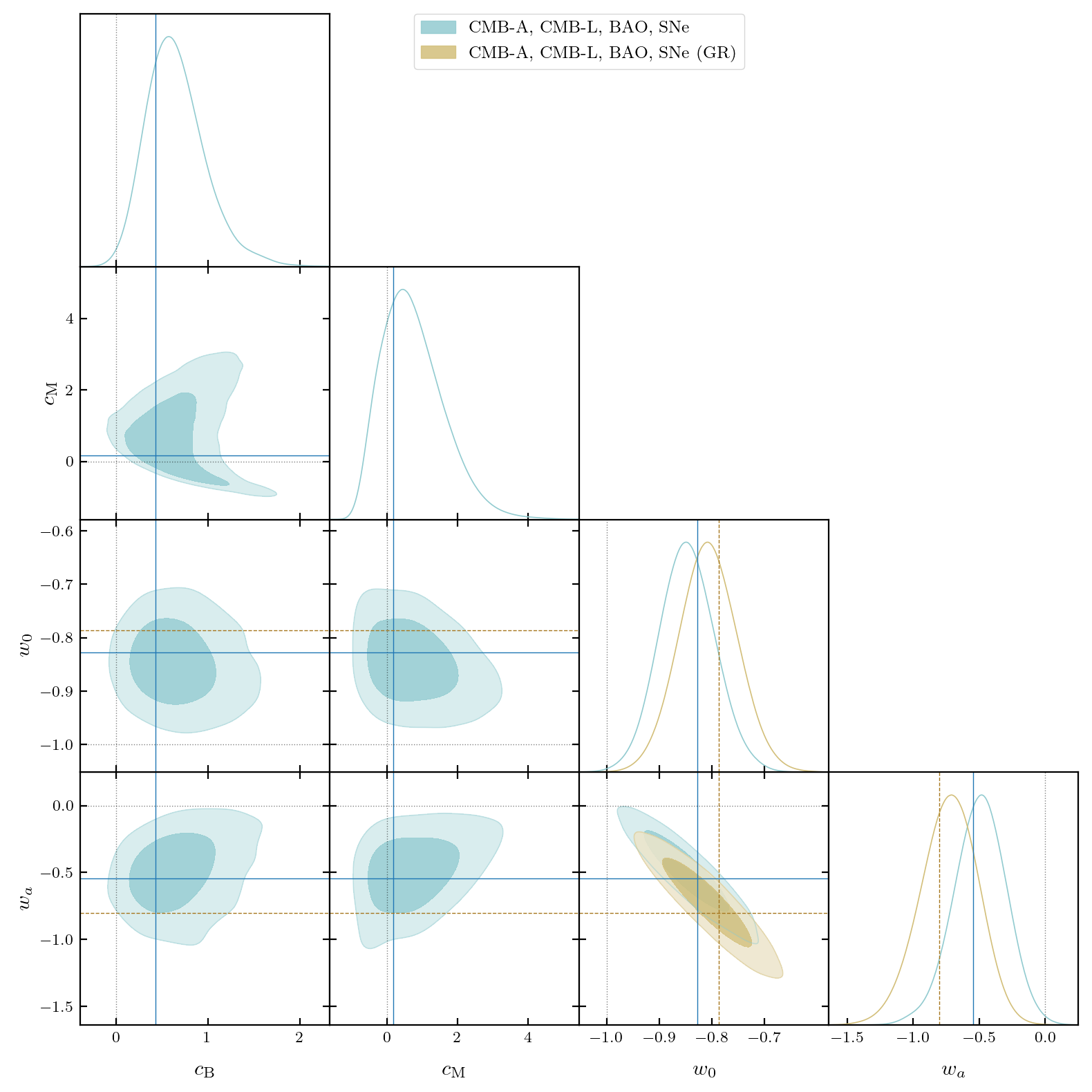}
    \caption{Joint posterior of $\{c_B, c_M, w_0, w_a\}$ for the EFTofDE model, and $\wzwa$ for $w_0w_a$CDM from our minimal dataset combination. The grey dotted lines represent the $\lcdm$ limits of the parameters, the yellow dashed line represents the MAP point for $w_0w_a$CDM, and the blue solid line represents the MAP point for EFTofDE. For the significance of deviation from $\lcdm$, on using the $\Delta\chi^2_{MAP}$ values at the MAP points for each model shown here and the number of extra degrees of freedom over $\lcdm$ (2 for $w_0w_a$CDM and 4 for EFTofDE), we get a 3.1$\sigma$ deviation for $w_0w_a$CDM and 2.9$\sigma$ deviation for EFTofDE.}
        
    \label{fig:minimal_rsd_isw_3x2pt_bm_w0wa}
\end{figure*}

We now ask the converse question of how constraints on the background expansion parameters $\wzwa$ are impacted by assumptions about DE perturbations. \cref{fig:backgroundonly_w0wa} shows the impact of the choice of model of DE perturbations on expansion history constraints. We compare the posteriors on $\wzwa$ in the EFTofDE model using the minimal combination with posteriors obtained with the same data combination in the standard so-called $w_0w_a$CDM model of dynamical DE, where subhorizon perturbations of DE are neglected and superhorizon perturbations are modelled with the Parametrised Post-Friedmann (PPF) approach \cite{Hu:2007pj,Fang:2008sn}. In the $w_0w_a$CDM model, gravity is described by GR. We find that a consistent modelling of subhorizon perturbations as employed by the EFTofDE brings the $\wzwa$ posteriors closer to $\lcdm$. As can be seen from \cref{fig:backgroundonly_bm}, the gradient stability prior prefers higher $\cBM$ values when the $\wzwa$ posterior is obtained with the early-universe CMB prior\footnote{This posterior was obtained for the EFTofDE model, but it is identical for the $w_0w_a$CDM model as this dataset combination is completely insensitive to DE perturbations.} This  $\wzwa$ posterior is shown in \cref{fig:backgroundonly_w0wa}, which is the $\wzwa$ projection of the same joint posterior that is projected in $\cBM$ in \cref{fig:backgroundonly_bm}. However, our probes of perturbations all constrain $\cBM$ to $\mathcal{O}(1)$ and disfavour large values. Through the gradient stability constraint on the joint space of $\{c_B, c_M, w_0, w_a\}$, the effect of this in the $\wzwa$ space is to disfavour large deviations from $\lcdm$, thus pulling the posterior closer to the $\lcdm$ expansion history.

\subsection{Assessing the significance of deviation away from $\lcdm$}

Dynamical dark energy modelled as $w_0w_a$CDM is known to deviate from $\Lambda$CDM at $\sim 3\sigma$ with the combination of Planck CMB, DESI DR2 BAO, and DES-Dovekie SNe observations at the time of this work \cite{DES:2025sig,Hoyt:2026fve}. It is interesting to ask whether a fundamentally consistent modelling of linear perturbations as done in the EFTofDE changes this significance of deviation \cite{Chudaykin:2024gol,Chudaykin:2025gdn,Lu:2025sjg} \togglered{(see e.g. \cite{Ye:2024ywg,Wolf:2025jed,Stolzner:2025vet,Ye:2026yqk,Garcia-Garcia:2026nzy} for calculations of observational evidence in other EFTofDE formulations different from the $\alpha$-basis used here.)} A naive expectation would be that the increase or decrease in significance is just dictated by how consistent the new extra parameters $\cBM$ remain with their GR limits. But that is not what we find: as demonstrated earlier in this section, the coupling between the background and perturbations due to the gradient stability condition shifts the background constraints towards the $\Lambda$CDM limit while shifting the perturbation constraints away from this limit. The impact of these competing effects on the significance of deviation from $\lcdm$ is not obvious and we do an explicit calculation to find the actual result.

There are minor variations in frequentist vs. Bayesian interpretations and dataset choices for CMB lensing in the literature. We perform this calculation for $w_0w_a$CDM for Planck PR4 lensing as the CMB lensing dataset (since this is what we use in our minimal dataset combination of CMB+BAO+SNe), and the deviation measure $\Delta\chi^2_{MAP}$, as used by the DESI collaboration\cite{DESI:2025zgx}. $\Delta\chi^2_{MAP}$ compares the values of $-2 \times$ the logarithm of the maximum a posteriori (MAP) points after an extra step of dividing the posterior by the prior volume added by uniform priors in the more complex model. This procedure retains the information in the MAP about any non-uniform priors (e.g. those for some CMB nuisance parameters) while removing penalties from arbitrary prior widths \cite{DESI:2025fii}.

We then compute the significance of deviation of the EFTofDE model from $\Lambda$CDM for the minimal dataset combination from the $\Delta\chi^2_{MAP}$ value between the EFTofDE and $\lcdm$ models by applying Wilk's theorem \cite{Wilks:1938dza}. The significance of deviation computed this way depends only on the value of $\Delta\chi^2_{MAP}$ and the number of extra degrees of freedom in the model compared to $\lcdm$. We show the significances for $w_0w_a$CDM and EFTofDE in \cref{tab:sigmadeviation}. We find that modelling DE perturbations with the EFTofDE results in only a very minor change in the deviation from $\lcdm$: 2.9$\sigma$ compared to 3.1$\sigma$ for $w_0w_a$CDM. This result is similar to the one found in \cite{Chudaykin:2024gol,Chudaykin:2025gdn,Lu:2025sjg} for a different combination of CMB+BAO+SNe datasets. It is quite nontrivial that the significance varies little between the two models: for example, the naive expectation from \cref{fig:backgroundonly_w0wa} would be lower significance of deviation for EFTofDE since the $\wzwa$ posterior moves closer to $\lcdm$ in EFTofDE, whereas the naive expectation from \crefrange{fig:adding_one_probe}{fig:adding_three_probes} would be a higher significance of deviation in EFTofDE since the $\cBM$ posteriors exclude $\lcdm$ at over 95\%. Ultimately, the lack of change is due to a compensation between several competing effects as we see in \cref{fig:minimal_rsd_isw_3x2pt_bm_w0wa} and \cref{tab:sigmadeviation}: 1) the posterior of $\wzwa$ being closer to the $\lcdm$ limit in EFTofDE than in $w_0w_a$CDM, 2) the posterior of $\cBM$ being away from their $\lcdm$ limit at more than 95\%, 3) the different extra degrees of freedom over $\lcdm$ in the two models (4 in EFTofDE and 2 in $w_0w_a$CDM), and 4) the difference in the maximum a posteriori values $\Delta\chi^2_{MAP}$ in the two models (-15.80 in EFTofDE and -12.57 in $w_0w_a$CDM).

\begin{table}[]
    \renewcommand{\arraystretch}{1.5}
    \centering
    \begin{tabular}{|c|c|c|c|}
    \hline
    \multirow{1}{*}{Model} & $\Delta\chi^2_{MAP}$ & \#dof & Significance \\
    \hhline{|=|=|=|=|}
    $w_0w_a$CDM & -12.57 & 2 & 3.1$\sigma$ \\
    \hline
    EFTofDE & -15.80 & 4 & 2.9$\sigma$ \\
    \hline
    \end{tabular}
    \caption{Comparison of the significances of deviation from $\lcdm$ for the $w_0w_a$CDM and EFTofDE models.}
    \label{tab:sigmadeviation}
\end{table}

\section{Conclusions} \label{sec:conclusions}
In this work, we simultaneously constrain the cosmic expansion history and linear cosmological perturbations in the EFTofDE framework of dynamical dark energy and modified gravity, using multiple probes of cosmic structure: redshift space distortions, weak gravitational lensing, and the Integrated Sachs-Wolfe effect, alongside state-of-the-art probes of cosmic expansion: CMB anisotropies and lensing, baryon acoustic oscillations, and Type IA supernovae. The vast suite of complementary datasets employed here allows us to understand the constraining power of each individual probe of structure in the EFTofDE parameter space comprising of $\cBM$ and interpret it physically with the help of the phenomenological functions $\{\mu(z), \Sigma(z)\}$. It also allows us to understand how consistent modelling at the Lagrangian level in the EFTofDE couples the posteriors of the background expansion $\wzwa$ and cosmological perturbations $\cBM$ through a theoretical prior coming from the requirement of positive sound speed or gradient stability. Our key findings are as follows:

\begin{itemize}

\item The minimal combination of probes that provides meaningful constraints on both the background expansion and perturbations in the EFTofDE consists of CMB anisotropies, CMB lensing, BAO and SNe. We improve the constraints on the DE perturbations described by the parameters $\cBM$ by supplementing this setup with various combinations of four probes of late-time cosmic structure: redshift space distortions, weak gravitational lensing measured through $\txtp$ correlations, and the Integrated Sachs-Wolfe effect measured through correlations of CMB temperature anisotropies with either galaxy number counts or CMB lensing. We explain our cautious approach of ensuring that covariances between probes are not neglected when they are expected to be present in \cref{sec:likelihood_combinations}. When considering each probe individually, in \cref{sec:adding_one_probe} we demonstrate their complementary by finding that that RSDs which solely probe clustering and are agnostic to lensing improve constraints on $c_M$, while $\txtp$, $\ITg$, and $\ITk$ which are stronger probes of lensing but weaker probes of clustering give stronger improvements on $c_B$, constraining a complementary direction in the parameter space.

\item We then combine multiple probes of structure with our minimal setup of CMB+BAO+SNe and demonstrate that combining complementary probes gives significant improvements in the constraints on $\cBM$ as shown in the posteriors in \cref{sec:multiprobe_constraints} and quantified by the Figure of Merit values in \cref{tab:eft_constraints}. Our most constraining probe combination where we can still ensure that there are no appreciable unmodelled covariances consists of the minimal setup of CMB+BAO+SNe supplemented with DESI DR1 RSD, DES Y3 $\txtp$, and ISW cross-correlations of Planck temperature anisotropies with galaxy number counts from photometric survey catalogs \cite{Stolzner:2017ged,Seraille:2024beb}. The strong constraints on the EFTofDE perturbations and complementarity of each constituent probe is demonstrated in \cref{fig:adding_three_probes}, and results in an improvement in the FoM for $\cBM$ by a factor of 2.69 compared to the $\cBM$ constraints from the minimal setup. 

\item We understand the phenomenological implications of the constraints on $\cBM$ by mapping them to the phenomenological functions $\{\mu(z), \Sigma(z)\}$ which directly affect the evolution of the gravitational potentials in \cref{sec:pheno_constraints}. We compute constraints on the present-day values $\mSt$ derived from each dataset combination, where we see stronger gains in FoM values compared to those seen for $\cBM$ (e.g. for the most constraining dataset combination shown in \cref{fig:minimal_rsd_isw_3x2pt_bm_w0wa,fig:minimal_rsd_isw_3x2pt_musigma}, the FoM improvement for $\mSt$ is by a factor of 3.37 while for $\cBM$ it is by a factor of 2.69) which is due to $\mSt$ having a more direct connection to observables. We also understand how the evolution of these functions over cosmic history is constrained by plotting $\{\mu(z), \Sigma(z)\}$ curves from samples that are within the 95\% posterior in the minimal dataset combination but are disfavoured by each new probe: RSD, $\txtp$, $\ITg$. From this, we conclude that RSD observations disfavour a large enhancement in the clustering of matter in the past, $\txtp$ observations similarly disfavour a large enhancement in lensing, and $\ITg$ measurements place constraints on the precise rate of decay of the Weyl potential $\Phi+\Psi$ at low redshifts $z\ \lsim\ 0.5$.

\item We demonstrate multiple ways in which constraints on the expansion history $\wzwa$ and perturbations $\cBM$ affect each other. First, in \cref{sec:background_choices} we show that switching the model for the background expansion history from $w_0w_a$ to a $\lcdm$-like expansion history ($w=-1$) has a substantial effect on the posteriors of $\cBM$, and surprisingly, for our minimal dataset combination, the constraints on $\cBM$ are weaker for the $\lcdm$ expansion history which has fewer phenomenological freedom in the background.

\item We show that the region of the $\cBM$ posterior close to the GR limit is affected not just by the constraining power of the data but also by the theoretical prior of gradient stability, which restricts the possible values of $\{c_B, c_M, w_0, w_a\}$ in the joint 4D parameter space. We quantify the effect of this prior on the posteriors of $\cBM$ conditional on the observed posteriors of $\wzwa$ by using the background-only dataset combination of the Planck early universe prior, BAO and SNe -- we find that in this setup where the entire posterior of $\cBM$ results from the gradient stability prior, the GR limit is disfavoured at more than 95\%. This is due to the observed expansion history deviating from a $\lcdm$ background, which forces $\cBM$ to be away from the GR limit.

\item Finally, we show how the observational question of the preference of dynamical DE over $\lcdm$ is affected by consistently modelling both the DE background and perturbations at the Lagrangian level in the EFTofDE, compared to the standard $w_0w_a$CDM formalism where phantom crossing values of $\wzwa$ are enables by ignoring subhorizon DE perturbations via the PPF formalism. We find the nontrivial result that the significance of deviation from $\lcdm$ for the more complex model changes only a little: 2.9$\sigma$ for EFTofDE compared to 3.1$\sigma$ for $w_0w_a$CDM.

\end{itemize}

There are multiple natural directions for extending this work. First, given the constraining power and complementarity of various LSS probes demonstrated in \cref{sec:combining_probes_constraints}, it is natural to work towards incorporating more measurements of various properties of LSS. Several combinations of LSS probes/datasets which are different from ours have been recently used in the literature to constrain the EFTofDE: for instance, KiDS-Legacy cosmic shear \cite{Stolzner:2025vet}; full-shape galaxy clustering from BOSS measured using EFTofLSS \cite{DAmico:2022osl} and photometric galaxies from the DESI Legacy Imaging Survey \cite{Zhou:2023gji} which were combined in \cite{Lu:2025gki,Lu:2025sjg}; and measurements of the decay rate of the Weyl potential \cite{Dong:2024vxo}, the lensing-inferred growth amplitude, and the growth rate from both DESI clustering and peculiar velocities, which were combined in \cite{Li:2026sbr}.

We compare our constraints on $c_B$ and $c_M$ with other recent works which constrain the EFTofDE with the same modelling choices as ours: namely, the $\alpha$-basis with $\cBM$ and the $w_0w_a$ expansion history. These works include the multiprobe analysis \cite{Lu:2025sjg}, the analysis with the first use of the Planck PR4 $\ITk$ likelihood to constrain the EFTofDE \cite{Chudaykin:2025gdn}, and the analysis with the first use of the DESI DR1 RSD+BAO likelihood to constrain the EFTofDE \cite{Ishak:2024jhs}. We compare the 1D uncertainties on $\cBM$ for our most constraining combination of minimal + RSD + $\txtp$ + $\ITg$ likelihoods as seen from \cref{tab:eft_constraints} with the tightest uncertainties on the same parameters found by \cite{Lu:2025sjg,Chudaykin:2025gdn,Ishak:2024jhs}. In our case, our tightest uncertainties are $(\Delta c_B, \Delta c_M) = (0.40, 0.97)$. These can be compared with $(\Delta c_B, \Delta c_M) = (0.38, 0.88)$ (\cite{Lu:2025sjg}), $(\Delta c_B, \Delta c_M) = (0.47, 1.41)$ (\cite{Chudaykin:2025gdn}), and $(\Delta c_B, \Delta c_M) = (0.44, 1.54)$ (\cite{Ishak:2024jhs}). Although the background expansion datasets are different across all these constraints (which have a non-negligible impact on $\cBM$ constraints as we demonstrate in \cref{sec:background_choices} and can also be seen in \cite{Chudaykin:2025gdn,Ishak:2024jhs} from the dependence of the constraints on the choice of SNe dataset), it is clear that the approach of combining complementary probes as pursued by us here and by \cite{Lu:2025sjg} gives substantially tighter constraints compared to adding only a probe of lensing or one of clustering to the minimal CMB+BAO+SNe combination.

Given this vast array of LSS probes now available, it is worthwhile to look for the optimal combination of these probes that avoids neglecting cross-probe covariances where they are expected to be present. On similar lines, it is also worthwhile to actually model the cross-probe correlations -- for instance, this has been done by \cite{Reeves:2025xau} where they modelled the cross-correlations between ACT DR6 CMB lensing \cite{AtacamaCosmologyTelescope:2025blo}, DESI photometric galaxies and Planck ISW, and by \cite{Lu:2025sjg} where they modelled the cross-correlations between Planck temperature and DESI photometric galaxies.

Secondly, on the theoretical front, it is important to constrain other EFTofDE models apart from the one we consider here, particularly models which have stronger fundamental motivations such as avoidance of instabilities in the presence of a gravitational wave background \cite{Creminelli:2019kjy}, or shift symmetry \cite{Pirtskhalava:2015nla,Traykova:2021hbr}. These models have been constrained in e.g. \cite{Noller:2020afd,Traykova:2021hbr,Shah:2025vnt}. Developing complementary priors is a further promising direction here, e.g. motivated by requiring radiative stability \cite{Noller:2018eht,Heisenberg:2020cyi}, somewhat related to the shift symmetry requirement discussed above, or from so-called positivity bounds on the underlying theory space, which arise as a consequence of demanding a sensible UV completion -- see \cite{Melville:2019wyy,Kennedy:2020ehn,deRham:2021fpu,Melville:2022ykg,deBoe:2024gpf} and references therein for how this can impact cosmological parameter constraints. 

Thirdly, comprehensively combining cosmological constraints with those derived from other scales holds great promising for  qualitatively improving our constraining power on the nature of dark energy further. Examples include local solar system constraints and their interplay with cosmology \cite{Babichev:2011iz,Burrage:2020jkj,Noller:2020lav}, current observational constraints from observed binary-black hole mergers and related to gravitational wave propagation \cite{Baka:2025drk,LIGOScientific:2026uyd,LIGOScientific:2026oim}, black hole ringdown constraints that take advantage of the `hair' generated by dynamical dark energy theories to constrain those further \cite{Noller:2019chl,deRham:2019gha,Khoury:2020aya,Takahashi:2021bml,Mukohyama:2022enj,Khoury:2022zor,DeFelice:2022xvq,Mukohyama:2023xyf,Sirera:2023pbs,Sirera:2024ghv,Mukohyama:2024pqe,Smulders:2026qwc,Smulders:2026bya},
as well as frameworks connecting a large range of astrophysical and cosmological scales 
\cite{Brax:2012gr,Baker:2014zba,Lombriser:2016zfz,Sanghai:2016tbi,Dima:2017pwp,Baker:2019gxo,Thomas:2020duj,Thomas:2025qiz}.

Finally, several LSS measurements also probe mildly nonlinear scales, which have the potential to considerably improve EFTofDE constraints as well as uncover new phenomenological degrees of freedom at higher orders in the EFTofDE Lagrangian. Making use of nonlinear scales requires development of novel theoretical predictions for the EFTofDE on nonlinear scales \cite{Kimura:2011dc,Takushima:2015iha,Cusin:2017mzw,Cusin:2017wjg,Amendola:2025xka,Sirera:2026klo} -- also see \cite{Thomas:2020duj,Srinivasan:2021gib,Fiorini:2022srj,Brando:2022gvg,Bose:2022vwi} for related approaches -- as well as incorporating these models in simulations \cite{Wright:2022krq,Brando:2023fzu,Ganjoo:2026ugf} so that analysis pipelines can be constructed.

\begingroup
\renewcommand{\addcontentsline}[3]{}

\section*{Acknowledgments}
\vspace{-0.1in}
\noindent 
We thank Seshadri Nadathur for insightful discussions. NS is supported by an UK Science and Technology Facilities Council (STFC) studentship (ST/X508688/1) and funding from the University of Portsmouth and University College London. KK is supported by STFC grant number ST/B001175/1. JN is supported by an STFC Ernest Rutherford Fellowship (ST/S004572/1). LY is supported by the National Key R $\&$ D Program of China (2023YFA1607800, 2023YFA1607803), NSFC grant 12525301, by the CAS Project for Young Scientists in Basic Research (No. YSBR-092), and by the Chinese Scholarship Council (CSC) and the University of Portsmouth.
In deriving the results of this work, we have used the following codes: hi\_class \cite{Zumalacarregui:2016pph,Bellini:2019syt}, Cobaya \cite{Torrado:2020dgo}, and GetDist \cite{Lewis:2019xzd}. We have made use of the SCIAMA HPC cluster at the Institute of Cosmology and Gravitation, University of Portsmouth, as well as the UCL Myriad High Performance Computing Facility (Myriad@UCL), and associated support services, in the completion of this work. For the purpose of open access, the authors have applied a Creative Commons Attribution (CC BY) licence to any Author Accepted Manuscript version arising from this work.
\\

\noindent{\bf Data availability} Supporting research data are available on reasonable request from the authors.

\endgroup

\appendix

\crefalias{section}{appendix}

\section{Further checks of the island in the $\ITg$ posteriors}\label{app-ISW}

\begin{figure}[t]
    \centering
    \includegraphics[width=\linewidth]{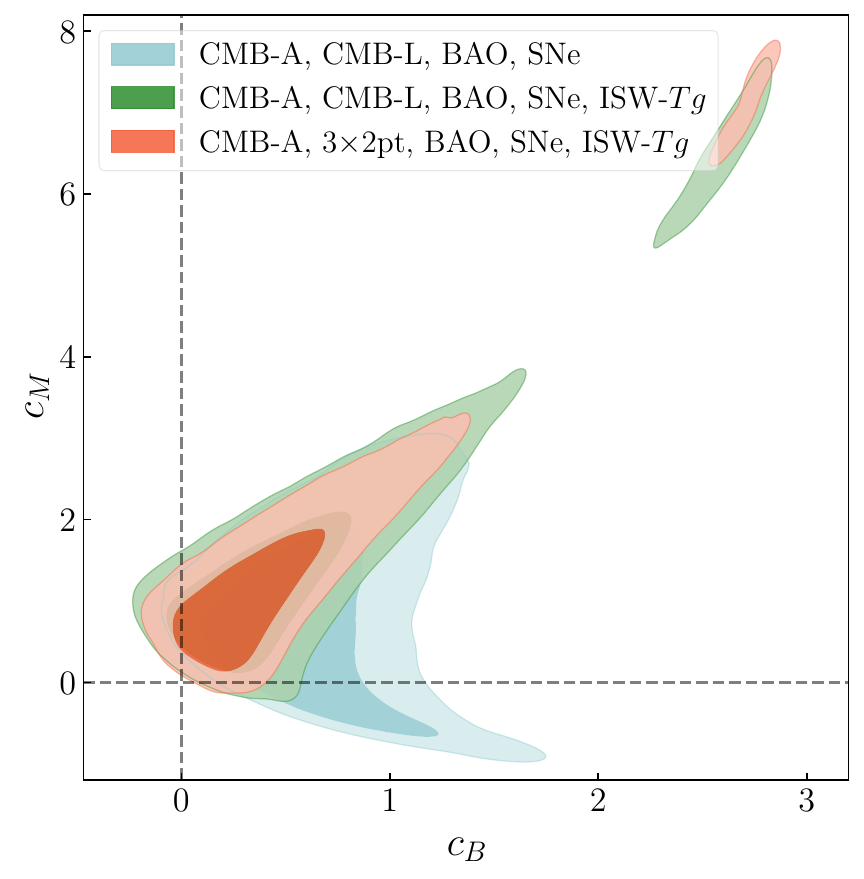}
    \caption{The posteriors on $\cBM$ for two dataset combinations including the $\ITg$ likelihood, which contain a small island within the 95\% region with large values of $\cBM$, separate from the main posterior peak. In \cref{fig:minimal_isw_island_predictions} we verify that despite the large $\cBM$ values in the island, the physical predictions for ISW cross-correlations in this part of the parameter space are reasonable.}
    \label{fig:withisland_comparison}
\end{figure}

\begin{figure}[t]
    \centering
    \includegraphics[width=\linewidth]{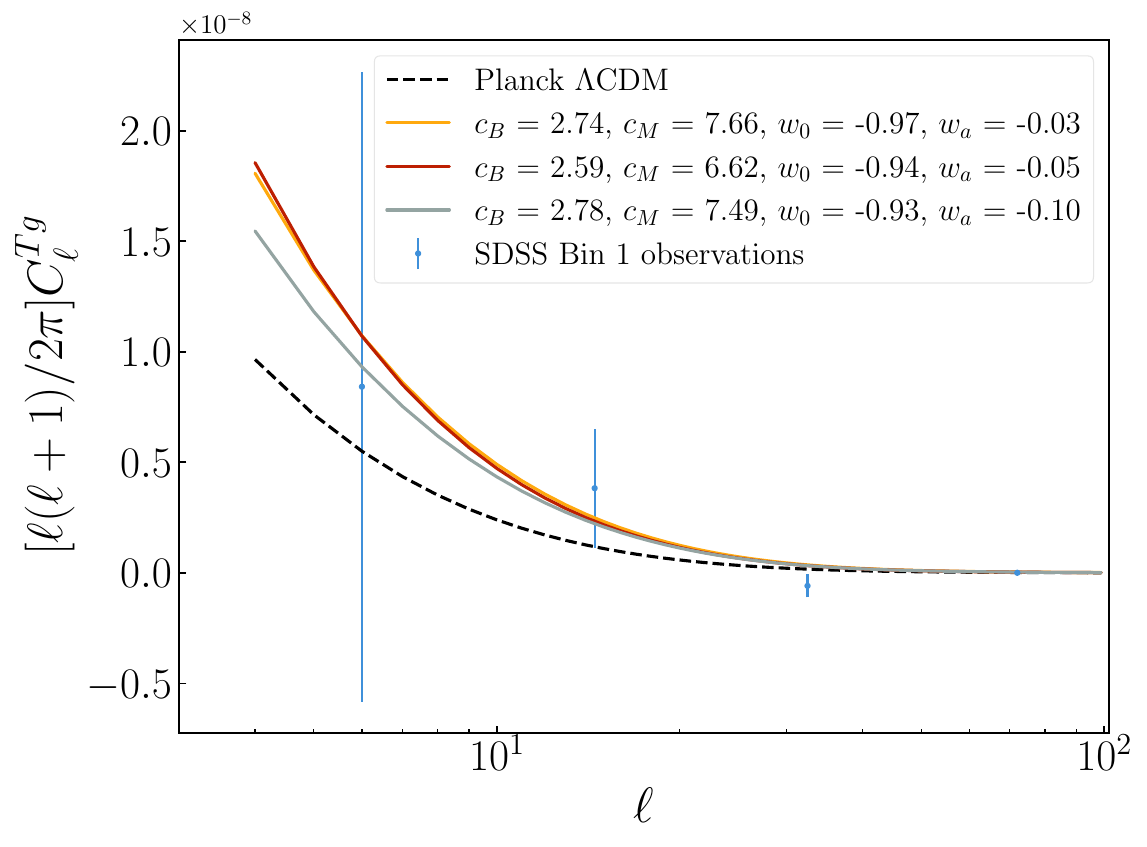}
    \caption{The ISW predictions for the island in the $\ITg$ posteriors shown in \cref{fig:withisland_comparison} in lowest redshift bin of the SDSS survey, with observations fron $z=0.1-0.3$. These predictions are calculated from samples from the chains which lie in the island. This plot shows that these points give reasonable physical predictions for ISW cross-correlations and fit the ISW signal in this bin comparably well as the Planck 2018 $\lcdm$ model.}
    \label{fig:minimal_isw_island_predictions}
\end{figure}

When using the $\ITg$ likelihood, for two specific dataset combinations our posteriors contain an island in the 95\% credible region, which shows up as a small second peak at much higher values of $\cBM$ as shown in \cref{fig:withisland_comparison}\footnote{We have sufficient convergence in the chains and have checked the numerical accuracy of our newly-implemented $\ITg$ and $\txtp$ likelihoods by comparing them with previous implementations in \cite{Shah:2025vnt,Wang:2024uvw}. This rules out unphysical causes for the island.}. This island is removed when RSD data is added to either of these data combinations. We confirm in \cref{fig:minimal_isw_island_predictions} that the physical predictions for the galaxy-CMB cross-correlations in this part of the parameter space are quite reasonable despite the unusually high values of $\cBM$.

\section{Impact of the $\Omega_m$ degeneracy}\label{app-Omegam}

\begin{figure}[htp!]
    \centering
    \includegraphics[width=\linewidth]{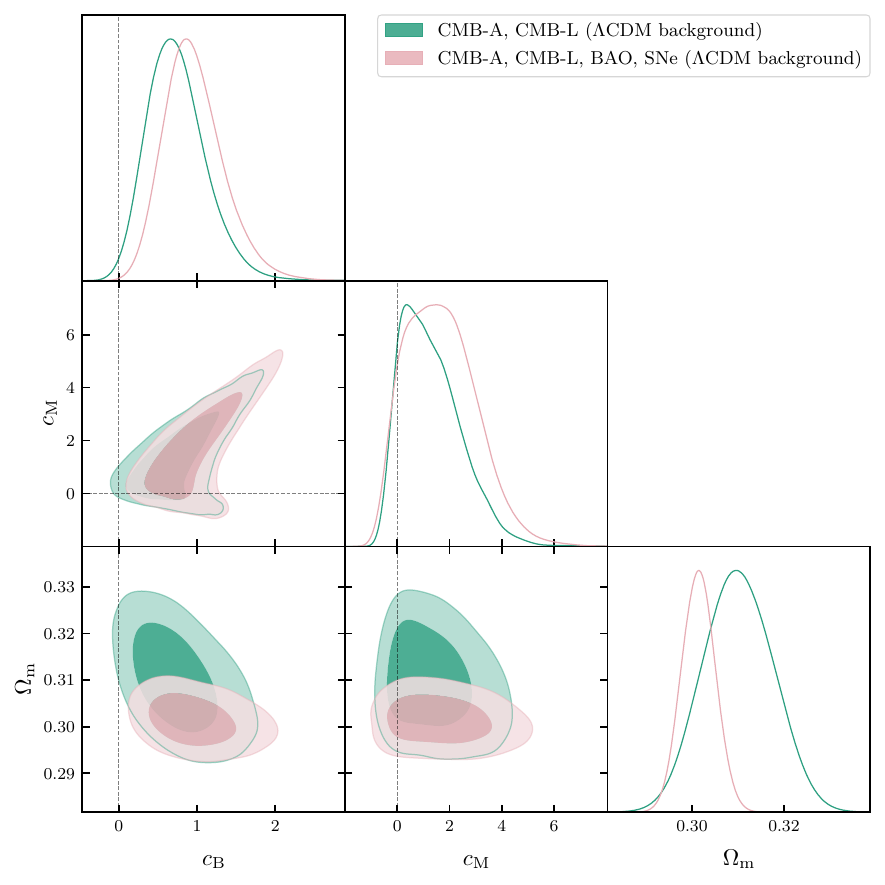}
    \caption{Comparison of posteriors on $\cBM$ and their joint posteriors with $\Omega_m$ assuming a $\lcdm$ expansion history with and without the addition of DESI DR2 BAO and DES Dovekie SNe datasets to Planck PR4 anisotropies and lensing. This plot demonstrates a negative degeneracy between $c_B$ and $\Omega_m$. This degeneracy combined with the difference in the inferred $\Omega_m$ values due to the impact of BAO and SNe datasets is responsible for the deviation from GR in $\cBM$ seen in the CMB+BAO+SNe dataset combination.}
    \label{fig:Omega_m_discrepancy}
\end{figure}

\cref{fig:Omega_m_discrepancy} shows that adding the BAO and SNe data to the CMB data significantly lowers the mean of the inferred $\Omega_m$ when a $\lcdm$ background is assumed. The underlying cause of this, as mentioned in \cref{sec:background_choices}, is the mild discrepancy in the values of $\Omega_m$ inferred in a $\lcdm$ model between the CMB and recent DESI BAO measurements \cite{DESI:2025zgx}. Due to the degeneracy between $c_B$ and $\Omega_m$ seen in their 2D posterior, this $\Omega_m$ discrepancy also increases the posterior mean of the inferred $c_B$, moving it away from the GR limit.

\bibliographystyle{utphys}
\bibliography{DESI_RSD_EFT}
\end{document}